\documentclass[11pt]{article}
\usepackage[margin=1in,top=.95in]{geometry}
\usepackage{amsmath, amssymb, braket,etoolbox,dsfont,authblk,titling}
\usepackage{graphicx}
\usepackage{rotating, mathtools}
\usepackage[matrix,arrow]{xy}
\usepackage[numbers]{natbib}
\usepackage[colorlinks]{hyperref}
\numberwithin{equation}{section}
\hypersetup{linkcolor={blue},citecolor={magenta}}

\hypersetup{linkcolor={blue}}
\newcommand{\A}[1]{O^{(#1)}}
\newcommand{\B}[1]{E^{(#1)}}

\newcommand{\QQ}[1]{Q^{(#1)}}
\newcommand{\ba}{\begin{align}}
\newcommand{\ea}{\end{align}}

\newcommand{\comment}[1]{}

\begin{document}

\title{Free fermions in disguise}%

\author{Paul Fendley\\
All Souls College and Rudolf Peierls Centre for Theoretical Physics,\\
Clarendon Laboratory, Parks Road, Oxford OX1 3PU 
}

\maketitle

\begin{abstract}
I solve a quantum chain whose Hamiltonian is comprised solely of local four-fermi operators by constructing free-fermion raising and lowering operators. The free-fermion operators are both non-local and highly non-linear in the local fermions. This construction yields the complete spectrum of the Hamiltonian and an associated classical transfer matrix. The spatially uniform system is gapless with dynamical critical exponent $z=3/2$, while staggering the couplings gives a more conventional free-fermion model with an Ising transition. The Hamiltonian is equivalent to that of a spin-1/2 chain with next-nearest-neighbour interactions, and has a supersymmetry generated by a sum of fermion trilinears. The supercharges are part of a large non-abelian symmetry algebra that results in exponentially large degeneracies. The model is integrable for either open or periodic boundary conditions but the free-fermion construction only works for the former, while for the latter the extended symmetry is broken and the degeneracies split.
\end{abstract}

\section{Introduction}

Free-fermion models yield profound insights in every area of theoretical physics. Essentially all such models are found by expressing the  Hamiltonian and/or action as a sum over bilinears of local fermionic operators or fields.  To put the Hamiltonian in this form, sometimes a non-local Jordan-Wigner transformation needs to be done \cite{Lieb61,SML}.  The purpose of this paper is to define and analyse a free-fermion model that requires a much subtler transformation, both non-local and non-linear in the original interacting fermions.  No form of the Hamiltonian as a sum over local bilinears of fermions seems to exist. 

A nice way of characterising a free-fermion model without having to delve into details of such transformations is via its spectrum. All eigenvalues $E$ of a free-fermion Hamiltonian are given in terms of energy ``levels'' $\epsilon_k$ as
\begin{align}
E=\pm \epsilon_1 \pm \epsilon_2 \pm \dots \pm \epsilon_S\ .
\label{ffE}
\end{align}
where each eigenvalue is specified by making a choice of each of the $\pm$ signs. The key property making the model free fermionic is that this choice does not affect the values of the $\epsilon_k$. All energies are given in terms of the same set of values $\epsilon_k$, which typically are found as the roots of an $S$th order polynomial.
Similarly, the eigenvalues of a transfer matrix for a classical free-fermion model are also given in terms of the $\epsilon_k$, only with a more complicated expression than \eqref{ffE}. 

The best way of solving a free-fermion model is to construct the raising and lowering operators. Advantages include not only making it possible to find the spectrum with boundary conditions other than periodic, but with spatially varying couplings as well. 
The raising and lowering operators $\Psi_{\pm k}$ obey
\begin{align}
\big[H,\,\Psi_{\pm k}\big]\ =\pm 2\epsilon_k\, \Psi_{\pm k}\ ,
\label{HPsi}
\end{align}
where $k=1,\dots, S$ labels the vector; in the periodic and uniform case one can instead label by the momentum. 
Acting with $\Psi_{\pm k}$ on an eigenstate of $H$ either annihilates the state, or gives a different eigenstate with energy shifted by $\pm 2\epsilon_k$. 
What makes these operators free fermionic is that they obey the anticommutation relations
\begin{align}
\big\{\Psi_{\pm k},\,\Psi_{\pm k'}\big\}=0\ ,\qquad\quad \big\{\Psi_{\pm k},\,\Psi_{\mp k'}\big\}= \delta_{kk'}\ .
\label{Psialg}
\end{align}
The operators $\Psi_k\Psi_{-k}$ for all $k$ then are commuting projectors:
\begin{align} 
(\Psi_k\Psi_{-k})^2 = \Psi_k\Psi_{-k}\ ,\qquad\quad \big[\Psi_k\Psi_{-k},\, \Psi_{k'}\Psi_{-k'}\big] =0 \ .
\label{commproj}
\end{align}
The fact that the spectrum is given by \eqref{ffE} then immediately follows. Since $\Psi_k$ and $\Psi_{-k}$ each square to zero and anticommute with each other, they can be thought of as creating or annihilating a fermionic particle of energy $\epsilon_k$ respectively. Moreover,
the Hamiltonian can be written as a sum over these bilinears as 
\begin{align}
H = \sum_{k=1}^S \epsilon_k^{} \big[ \Psi_{k},\,\Psi_{-k} \big]\ ,
\label{He}
\end{align}
as suggested by (\ref{HPsi}) and \eqref{commproj}.

When the Hamiltonian is a bilinear in free-fermionic operators, constructing the raising and lowering operators is straightforward. Because commuting a fermion bilinear with any operator linear in fermions gives a fermion linear back again, the commutator then can be represented as a matrix acting on the space of fermion linears. The rank of the matrix grows only linearly with the size of the system, and varying the couplings simply varies the entries. The eigenvectors determine the raising and lowering operators, while the eigenvalues give the $\epsilon_k$, as clear from \eqref{HPsi}. 
This procedure works for a host of interesting models, and by now, the technology for doing this construction is standard and simple.

In this paper, I describe and solve a model that is free-fermion as described above, but where the standard construction does not apply.  No Jordan-Wigner-type transformation gives a local Hamiltonian bilinear in fermionic operators. Even with spatially uniform couplings, the Hamiltonian is purely-four fermion: 
\begin{align}
H_{\rm u}= \sum_{j=1}^{2M}\,\psi^{}_{j}\psi^{}_{j+1}\psi^{}_{j+3}\psi^{}_{j+4}\ ,
\label{Hfu}
\end{align}
where the operators $\psi_j$ with $j=1\dots 2(M+2)$ obey the usual Majorana-fermion Clifford algebra
\begin{align}
\big\{\psi_j,\,\psi_{j'}\}= 2 \delta_{jj'}\ .
\label{Majalg}
\end{align}
The equivalent spin Hamiltonian is given in \eqref{hh} and \eqref{HH} below.  Despite its having no local fermion-bilinear form, with open boundary conditions this Hamiltonian has a spectrum given by \eqref{ffE}, along with raising and lowering operators that satisfy all the free-fermion properties \eqref{HPsi}--\eqref{He}. As befitting a free-fermion model, the construction still works when taking an arbitrary coupling in front of each term in \eqref{Hfu}. It also works for a family of commuting classical transfer matrices, whose spectra are determined in terms of the same $\epsilon_k$.

Although the construction given here seem rather miraculous, it is inspired by the solution of a ${\mathbb Z}_n$-invariant Hamiltonian of Baxter's \cite{Baxter89a,Baxter89b}. Exact raising/lowering operators can be constructed for it using ``free parafermions'', yielding the entire energy spectrum of the original Hamiltonian \cite{Fendley14} and an associated transfer matrix \cite{Fendley14,Baxter14,AuYang14} with open boundary conditions. The free-parafermion construction utilises only the algebra of the generators of the Hamiltonian and transfer matrix, which in this case is quite simple.  Unfortunately, the Baxter Hamiltonian is not Hermitian and its spectrum is complex, except in the $n=2$ case where it reduces to Ising. However, the beautiful algebraic way in which it is solved suggests that the method will apply to other models constructed from generators obeying a simple algebra. The main purpose of this paper is to show that \eqref{Hfu} is a hermitian Hamiltonian indeed solvable in the same fashion.

In addition to this marvellous property, the Hamiltonian \eqref{Hfu} is very interesting in its own right. It is a Majorana version \cite{Sannomiya17b} of a model introduced long ago by Nicolai as simple example of lattice supersymmetry \cite{Nicolai76}. It reappeared as a limiting case of a self-dual Hamiltonian introduced to analyse the physics of the tricritical Ising Hamiltonian and the appearance of supersymmetry there \cite{OBrien17}. Indeed, the uniform Hamiltonian \eqref{Hfu} can be written as 
\begin{align}
H_{\rm u} = \frac{1}{2}\mathcal{Q}^2 - 2M ,\qquad\hbox{ where }
\mathcal{Q}= \sum_{j=1}^{2M} \psi_{j}\psi_{j+1}\psi_{j+2}\ ,
\label{HQ}
\end{align}
so that supercharge $\mathcal{Q}$ commutes with $H_{\rm u}$. 
It is worth noting that a supercharge trilinear in fermions also appears in the zero-dimensional supersymmetric SYK model \cite{Fu16}. 

Not only is \eqref{Hfu} supersymmetric, but I show below that the supersymmetry algebra extends to a much larger non-Abelian one commuting with the Hamiltonian/transfer matrix. 
However, the generators of this large symmetry algebra do not commute with the raising and lowering operators. Because their number increases linearly with $M$, each energy level has a degeneracy growing exponentially with $M$. In both this respect and the fact that the spectrum is free-fermionic, the model analysed here is reminiscent of the ``Cooper pair'' supersymmetric chain of \cite{Fendley06}. Another similarity is that the degeneracies in both models are broken by taking periodic boundary conditions. An important difference though is that the degeneracies there are much subtler, coming from Cooper pairs whose number depends on the level. Here, all levels have the same degeneracy. Moreover, the $U(1)$ symmetries of the Cooper-pair chain make it simple to solve using the Bethe ansatz because of the presence of $U(1)$ symmetries. While the model here is integrable with both open and periodic boundary conditions, the lack of a $U(1)$ symmetry makes the free-fermion approach described here much more suitable.

A striking property derived here is that \eqref{Hfu} is critical but not Lorentz-invariant in the continuum limit. Instead, the excitations have a dispersion relation with dynamical critical exponent $z=3/2$. When the couplings are staggered with periodicity of three sites, the system exhibits three critical lines with $z=1$ meeting at the uniform point \eqref{Hfu}. The critical lines are of a more conventional free-fermion/Ising type, with spontaneously broken supersymmetry. The $z=3/2$ uniform point is thus a novel multicritical point separating three phases with different ordering, as shown in the phase diagram in figure \ref{fig:phasediagram} below. 

In section \ref{sec:hamctm}, I introduce the model in more detail, and find an extensive set of conserved quantities for the Hamiltonian and a more general transfer matrix.  The open chain is solved in section \ref{sec:open} by explicit construction of the raising and lowering operators. The generators of the supersymmetry and extended algebras are found in section \ref{sec:susy}. The spectrum and phase diagram for the uniform and staggered models is explicitly derived in section \ref{sec:staggered}.  I collect some technical calculations in the appendices, in particular finding an $R$ matrix that shows that the models are integrable in the traditional sense for periodic or open boundary conditions.

\section{Conserved charges and commuting transfer matrices}
\label{sec:hamctm}

\subsection{The Hamiltonian and the generating algebra}
\label{sec:ham}

The operators in the quantum and classical models studied in this paper act on a Hilbert space constructed from two-state quantum ``spin'' systems arranged into a chain. A useful basis of such operators acting on this space is comprised of
\begin{align}\sigma^a_m\equiv 1\otimes 1\cdots 1\otimes \sigma^a \otimes 1\cdots 1\ ,
\label{sigmaam}
\end{align}
where the Pauli matrix $\sigma^a$ acts on the two-state system labelled by $m$. The Hamiltonians and transfer matrices of central interest here are built from the operators 
\begin{align}
h_m = b^{}_m\,\sigma^z_m\sigma^z_{m+1}\sigma^x_{m+2}\ , \quad\qquad \widetilde{h}_m = \widetilde{b}^{}_m\, \sigma^x_m\sigma^z_{m+1}\sigma^z_{m+2}\ ,
\label{hh}
\end{align}
where the ${b_m}$ and $\widetilde{b}_m$ are real parameters. Open boundary conditions amount to taking the number of two-state systems to be $M+2$, so that the Hilbert space is of dimension $2^{M+2}$, whereas closed correspond to identifying  $\sigma^a_{m+M}\equiv\sigma^a_m$ (and hence $h_{m+M}\equiv h_m$), so that the dimension is $2^M$. The Hamiltonians studied in this paper are then
\begin{align}
H=\sum_{m=1}^M \,h_m\ ,\quad\qquad \widetilde{H}= \sum_{m=1}^M\widetilde{h}_m\ ,
\label{HH}
\end{align}
The different types of operators defined in \eqref{hh} have the very useful property that they commute with each other:
\begin{align}
\big[h_m,\, \widetilde{h}_{m'}\big] = 0\quad \hbox{for all }m\qquad\Rightarrow \big[H,\,\widetilde{H}\big] =0\ .
\end{align}
The two thus can be diagonalised independently.  Since with an appropriate relation of the $b_m$ and $\widetilde{b}_m$, $H$ and  $\widetilde{H}$ are parity conjugates of each other, it suffices to analyse one of them.

The free-parafermion solution \cite{Fendley14} of Baxter's $\mathbb{Z}_n$ Hamiltonian \cite{Baxter89a,Baxter89b} relies crucially on its generators satisfying a simple algebra. The generators here satisfy a simple algebra as well:
\begin{align}
\nonumber
h_m^2 = (b_m)^2 ,\qquad  &h_m h_{m+1}=-h_{m+1}h_m\ ,\qquad
  h_m h_{m+2}=-h_{m+2}h_m\ , \\  &h_m h_n = h_n h_m\ \hbox{ for }\ |n-m|>2\ .
\label{halg}
\end{align}
for $m=1\dots M$.  The generators $\widetilde{h}_m$ satisfy the same algebra. For periodic boundary conditions, these relations are interpreted with indices mod $M$, while for open they are not; e.g. $h_{M-1}$ and $h_1$ anticommute for periodic boundary conditions but commute for open. I will show how the exact solution for open boundary conditions requires only utilising properties of a slightly extended version of the algebra \eqref{halg}, without recourse to the explicit representation \eqref{hh}. This analysis will then apply to any model with Hamiltonian $H$ whose generators obey \eqref{halg}, such as the model with Hilbert-space dimension $\sim 2^{M/2}$ given in section \ref{sec:newham} below. In the special case $b_{3j}=0$, the remaining $h_{m}$ satisfy the same algebra as do the generators of the Ising Hamiltonian, and the analysis here then simplifies to the standard construction of raising and lowering operators for Ising (see e.g.\ \cite{Fendley14}).

The four-fermion Hamiltonian \eqref{Hfu} corresponds to $H_{\rm u}=H+\widetilde{H}$ with all $b_m=\widetilde{b}_m=1$, a simple consequence of the Jordan-Wigner transformation
\begin{align}
\psi^{}_{2m-1}=\sigma^z_m\prod_{n=1}^{m-1}\sigma^x_n\ ,\qquad\quad
\psi^{}_{2m}=-i\sigma^x_m\psi_{2m-1}\ .
\label{JW}
\end{align}
Although $H_{\rm u}$ is a special case of a Majorana-fermion analog \cite{Sannomiya17b,OBrien17} of a supersymmetric model introduced long ago \cite{Nicolai76}, the decomposition into two commuting models does not seem to have been observed before. 
Its physics is quite different from the pure four-fermion Hamiltonian comprised of terms with consecutive operators $\psi_m\psi_{m+1}\psi_{m+2}\psi_{m+3}$ studied in \cite{Selke88,Rahmani15a}. For example, the latter is gapped, while $H_{\rm u}$ is gapless, as shown in section \ref{sec:staggered}. More insight into the distinction between these two models will appear soon \cite{AA19}.

\subsection{Conserved charges}

Expressing the Hamiltonian in terms of generators satisfying (\ref{halg}) makes it easy to show that the Hamiltonian commutes with an extensive number of conserved charges. The method for constructing them is quite similar to that used for free parafermions \cite{Fendley14}. Non-local conserved charges are constructed in terms of products of different $h_m$ that commute with each other, e.g.\ $h_m\,h_{m'}$ with $|m-m'|>2$ mod $M$.  For example, the sum 
\[
\QQ{2} \equiv \sum_{|m-m'|>2} h_m\,h_{m'}
\]
commutes with $H$, as is easy to verify because $\QQ{2}=H^2/2$ plus a constant. An entire hierarchy of conserved charges is defined by the same rule: 
\begin{align}
\QQ{s} \equiv \sum_{\{M\ge m_{r+1}>m_{r}+2\}}
h_{m_1}\,h_{m_2}\,\dots\,h_{m_s}\ ,
\label{Jdef}
\end{align}
where $\QQ{1}=H$. The notation in the summation means to sum over all $m_r=1\dots M$ with $r=1\dots s$ subject to the constraints $m_{r+1}>m_r+2$, with the additional constraint $m_{s}-m_1\ne M-1,M-2$ for periodic boundary conditions. To prove they commute with the Hamiltonian for all $s$, first consider commuting $H$ with a single term in $\QQ{s}$:
\[ \big[H, h_{m_1}\,h_{m_2}\,\dots\,h_{m_s}\big] 
= \sum_{m=1}^M \big[h_m, h_{m_1}\,h_{m_2}\,\dots\,h_{m_s}\big]=
2\sum_{\substack{|m-m_r|=1,2\\ |m-m_{r\pm 1}|\ne1,2 }} h_m h_{m_1}\,h_{m_2}\,\dots\,h_{m_s}\ ,\]
by using the algebra \eqref{halg}.
The latter sum is over all $m$ such that $m$ is one or two sites from a single but not two $m_r$. The latter restriction arises because e.g.\  $[h_{m+1}, h_{m} h_{m+3}]=0$. It means there always occurs another contribution to $[H,\,\QQ{s}]$ where the indices $m$ and $m_r$ change places: $h_m$ comes from a term in $\QQ{s}$, while $h_{m_{r}}$ comes from $H$. These pairs cancel in the sum over the $m_r$:
\[ [h_m,\, h_{m_1}\dots h_{m_{r-1}} h_{m_r}h_{m_{r+1}}\dots \,h_{m_s}] + [h_{m_r},\, h_{m_1}\dots h_{m_{r-1}}h_{m}h_{m_{r+1}}\dots h_{m_s}] = 0\ ,\]
yielding $[H,\,\QQ{s}]=0$.

When given a hierarchy of commuting conserved charges, it is natural to define the {\em transfer matrix} of a two-dimensional classical lattice model that commutes with the quantum Hamiltonian.  It is simply the generating function of these charges
\begin{align}
T_M(u) = \sum_{s=0}^{S} (-u)^s Q^{(s)} \ ,
\label{TJdef}
\end{align}
where $u$ is a real parameter. The sum truncates at $S=[M/3]$ for periodic boundary conditions and $S=[(M+2)/3]$ for open, with $[x]$ the integer part of $x$. For open boundary conditions, a very useful equivalent definition of the transfer matrices are via the recursion relation
\begin{align}
T_{M}(u)=T_{M-1}(u)-uh_MT_{M-3}(u)\ .
\label{Trecur}
\end{align}
for $M\ge 1$, with $T_M=1$ for $M\le 0$.  A more traditional expression of this transfer matrix in terms of a vertex model (i.e.\ a matrix product operator) is given in Appendix \ref{app:Rmat}. The Boltzmann weights are local, but I am not aware of any previous appearance in the literature.

In the appendices I present two proofs that not only do the conserved charges $\QQ{s}$ commute with the Hamiltonian, but also with one another, so that
\begin{align}
\big[ T_M(u),\,T_M(u')\big]= 0\ 
\label{Tcommute}
\end{align}
for all $u$ and $u'$. 
The parameter $u$ is typically called the {\em spectral parameter}, since whereas the spectrum of $T$ depends on it, the eigenvectors do not.
In appendix \ref{app:Rmat}, I give a proof of \eqref{Tcommute} valid for both open and periodic boundary conditions by using the standard Yang-Baxter technique \cite{Baxter82}.  I supply an alternative proof in Appendix \ref{app:CTMprod} using the product form derived next. The latter proof has the disadvantage that it applies only for open boundary conditions, but the advantage that it requires only use of the algebra (\ref{halg}). It thus applies to any Hamiltonian and transfer matrix whose generators obey \eqref{halg} such as that given in section \ref{sec:newham}, not only those of the form (\ref{hh}).

\subsection{The product form and inverse of the transfer matrix}

The main result of this paper is that for open boundary conditions, $H$ and $T_M(u)$ can be rewritten in terms of non-local free-fermion operators.  Despite its being integrable both for periodic and open boundary conditions, its features are much more striking in the latter case. Indeed, it appears the free-fermion solution only applies for open boundary conditions. Henceforth the analysis in this paper applies to having open boundary conditions. 

The method for deriving this remarkable property relies on rewriting the transfer matrix as a product of local (but not commuting) operators. Namely,
\begin{align}
T_M(u) = G_{M}(u) G_{M}^{\rm T}(u)\ ,\qquad G_{M}(u)\equiv g_1g_2\dots g_{M}\ .
\label{Gdef}
\end{align}
where the superscript ${\rm T}$ means transpose, with $g^{\rm T}_m=g_m$. The basic building blocks of this product form are the local operators
\begin{align}
g_m = \cos\frac{\phi_m}{2} + \frac{h_m}{ b_m}\sin\frac{\phi_m}{2}\ ,
\end{align}
where the angles $\phi_m$ are defined recursively via
\begin{align}
\sin\phi_{m+1}&= -\frac{ub_{m+1}}{\cos\phi_{m-1}\cos\phi_{m}}\ ,
\label{phidef}
\end{align}
with $\phi_0=\phi_{-1}=0$. The spectral parameter $u$ can always be chosen small enough to make all the $\phi_m$ real and the $g_m$ hermitian.   
These operators obey $g_mg_{m'}=g_{m'}g_m$ when $|m-m'|>2$, while 
\begin{align}
g_{m} h_{n} g_m &=  h_{n} \cos\phi_m\qquad \hbox{for }|n-m|=1\hbox{ or } 2\ .
\label{gident1}
\end{align}

The proof of \eqref{Gdef} follows by induction using the recursion relation (\ref{Trecur}), along with the identity \eqref{gident1} and
\begin{align}
(g_m)^2 -1 &= \frac{h_m}{ b_m} \sin\phi_m=-h_m\frac{u}{\cos\phi_{m-1}\,\cos\phi_{m-2}}\ .
\label{gident}
\end{align}
Assuming (\ref{Gdef}) holds up to some given $M-1$ implies
\begin{align*}
G_{M}G^{\rm T}_{M}= G_{M-1}g^2_M G^{\rm T}_{M-1}&=T_{M-1}+G_{M-1}(g^2_M-1) G^{\rm T}_{M-1}\cr
&=T_{M-1}+G_{M-3}(g^2_M-1)\cos\phi_{M-1}\cos\phi_{M-2} \,G^{\rm T}_{M-3}\cr
&=T_{M-1}-G_{M-3}uh_M\,G^{\rm T}_{M-3}\cr
&=T_{M-1}-uh_M T_{M-3}=T_{M}\ ,
\end{align*}
since $h_M$ and hence $g_M$ commute with $G_{M-3}$.
Since  $G_{-2}=G_{-1}=G_0=1$, by induction (\ref{Gdef}) therefore holds for all $M$.

The product form (\ref{Gdef}) makes finding the inverse of $T(u)$ easy. 
Since sending $u\to -u$ sends all angles $\phi_m\to -\phi_m$, it follows immediately that $g_m(\pm u)g_m(\mp u) = \cos\phi_m$ and
\begin{align}
&G_{M}(-u)G_{M}^{\rm T}(u) = G^{\rm T}_{M}(-u)G_{M}(u) = \prod_{m=1}^{M} \cos\phi_m\ ,\cr
&T_M(u)T_M(-u) = \prod_{m=1}^{M} \cos^2\phi_m\equiv P_M^{}(u^2)\ 
\label{GTinv}
\end{align}
Thus the transfer matrix is invertible except when $u$ is a root of $P_M(u^2)$, and when it exists, the inverse of $T_M(u)$ is proportional to $T_M(-u)$. 

The function $P_M(u^2)$ is a polynomial in $u^2$, as follows from the 
recursion relation
\begin{align}
P_{m}^{}(u^2)=P_{m-1}^{}(u^2)-u^2b_m^{2}\,P_{m-3}^{}(u^2)
\label{Precur}
\end{align}
for $m>0$, with $P_0=P_{-1}=P_{-2}=1$. This relation follows simply from the recursive definition  \eqref{phidef} of the $\phi_m$:
\[\frac{P_{m}}{P_{m-1}}=\cos^2\phi_{m}=1-\sin^2\phi_{m}=1-u^2b_{m}^2\frac{P_{m-3}}{P_{m-1}}\ .\]
The polynomial $P_M(u^2)$ is thus of order $S\equiv[M/3]$, where $[x]$ means the integer part. The similarity of the recursion relations \eqref{precur} and (\ref{Trecur}) means that the polynomial takes on the same form as the transfer matrix, with the role of the $h_m u$ in the latter replaced by $(b_m u)^2$ in the former. Namely, its coefficients are sums of products of the $b_m^2$, such that in any product $b_mb_{m'}$ appears only if $|m-m'|>2$. For example,
\begin{align*}
P_1&=1-b_1^2u^2\ ,\quad P_2=1-(b_1^2+b_2^2) u^2\ ,\quad P_3=1-(b_1^2+b_2^2+b_3^2) u^2\ ,\cr 
P_4&=1-(b_1^2+b_2^2+b_3^2+b_4^2) u^2 + b_1^2b_4^2 u^4\ .
\end{align*}
This correspondence between transfer matrix and polynomial holds in the Ising and free parafermion models with open boundary conditions as well \cite{Fendley14}. With the normalization defined here, the polynomial can be written in terms of its roots $u_k^2$ as 
\begin{align}
P_{M}(u^2) = \prod_{k=1}^S\left(1-\frac{u^2}{u_k^2}\right)\ .
\label{Puk}
\end{align}

\subsection{Higher Hamiltonians}
\label{sec:higherH}

The fact the non-local conserved charges all commute with each other makes possible the construction of conserved charges that are the sum of local operators. These ``higher Hamiltonians'' are generated by the logarithmic derivative of $T_M$:
\begin{align}
\sum_{r=1}^{\infty} H^{(r)}u^{r-1} \equiv {\cal H}(u)\equiv - \frac{d}{du}\ln T(u)=- \frac{1}{P_M(u^2)}T_M(-u)T_M'(u)\ ,
\label{higherH}
\end{align}
where the prime here means $d/du$. 
This relation should be understood as a formal definition of the higher Hamiltonians $H^{(r)}$ given by expanding the final expression in a series in $u$ around $u=0$. Equivalently, one can exploit analyticity. 
Since the order $u^0$ term in the polynomial $P_M(u^2)$ is non-vanishing, none of its roots are at $u=0$. Therefore ${\cal H}(u)$ is analytic in a region around $u=0$,  since $T(u)$ is a finite series in $u$ with coefficients given by the bounded (for finite $M$) operators $Q^{(s)}$. Moreover, ${\cal H}(u)$ is a meromorphic function in $u$, since the only singularities arise at the zeroes of $P_M(u^2)$. Thus the higher Hamiltonians can be extracted by contour integrals around $u=0$: 
\begin{align}
H^{(r)} =\frac{1}{2\pi i}\oint du\, u^{-r} {\cal H}(u)\ 
\label{Hcontour}
\end{align}
for $r$ a positive integer.


This definition (\ref{higherH}) of course yields $H^{(1)}=Q_M^{(1)}=H$. The Hamiltonians $H^{(2s)}$ are simply constants, as is easy to show from taking the derivative of the inversion relation (\ref{GTinv}) for $T_M(u)$:
\begin{align}  P'_M(u^2)= T_M(-u)T_M'(u) - T_M'(-u)T_M(u)
= -2P_M(u^2)\sum_{s=1}^\infty  H^{(2s)} u^{2s-1}\ .
\label{Pprime}
\end{align}
Therefore all even Hamiltonians are diagonal operators, e.g.\
\begin{align*}
H^{(2)}=\sum_m b_m^2\  ,\qquad\ 
H^{(4)}=\sum_m b_m^2(b_m^2+2b_{m+1}^2+2b_{m+2}^2) \ .
\end{align*}
Writing $P_M(u^2)$ in terms of its roots as in \eqref{Puk} and expanding out the product in $P'_M/P_M$ gives simple expressions for all even Hamiltonians in terms of the zeroes:
\begin{align}
\frac{P'_M(u^2)}{P_M(u^2)}=(\ln P_M)' =-\sum_{k=1}^S \frac{2u}{1-u^2/u_k^2}\ \qquad\Rightarrow\quad
H^{(2s)} = \sum_{k=1}^S u_k^{-2s}\ .
\label{NewtonGirard}
\end{align}

\section{The solution of the open chain}
\label{sec:open}

\subsection{The spectrum and the degeneracies}
\label{sec:spectrum}

The product form of the transfer matrix with open boundary conditions makes it possible to find the raising and lowering operators explicitly, and derive many interesting properties.  Because there is no way to rewrite the Hamiltonian in terms of local free-fermion bilinears, the analysis is much more involved than typical for free-fermion models (although it is essentially all straightforward algebra). I thus summarise the results of this section here.

The basic properties of the raising and lowering operators are summarised in the Introduction. In this section I construct these operators explicitly , and show how the energy levels $\epsilon_k$ are related to the $S$ roots $u_k^2$ of the polynomial $P_M(u^2)$ defined in (\ref{Precur}). Namely, 
\begin{align}
\epsilon_{k} =\frac{1}{u_k}\ ,\qquad\hbox{ for } k=1\dots S\ .
\label{epsilonu}
\end{align}
There are thus $S=[(M+2)/3]$ pairs of raising and lowering operators. Since $H$ is Hermitian and the $\epsilon_k$ are the coefficients in its commutator with the $\Psi_k$, they must be real. Thus even though it is not obvious from the definition, all the roots $u_k^2$ must always be positive and real, so $\epsilon_k$ is also taken positive and real.

Because they obey (\ref{HPsi}), acting with a raising and lowering operator on an eigenstate of $H$ either annihilates the state, or gives an eigenstate with energy shifted by $\epsilon_k$. Moreover, because $\{\Psi_k,\Psi_{-k}\}=1$ and $\Psi^2_k=\Psi_{-k}^2=0$, all eigenstates can be grouped into pairs where one state is annihilated by $\Psi_k$ and the other by $\Psi_{-k}$, the former with energy $2\epsilon_k$ higher than the latter. The projection operator $\Psi_{k}\Psi_{-k}$ therefore projects onto the former, while $\Psi_{-k}\Psi_{k}$ projects onto the latter. These projection operators with $k=1\dots S$ all commute with each other, so these pairs then can be grouped into multiplets of dimension $2^S$. Since the Hamiltonian in the form \eqref{He} is a sum over these projection operators each such multiplet has energies 
\begin{align}
E= \pm \frac{1}{u_1} \pm \frac{1}{u_1} \pm \dots \pm  \frac{1}{u_S} \ .
\label{EEu}
\end{align}
Each $\pm$ sign can be chosen differently, giving $2^S$ distinct energies for the states in the multiplet. This spectrum is precisely that of a free-fermion theory, i.e.\ precisely of the form (\ref{ffE}) with $\epsilon_k=1/u_k$. 
The spectrum of the higher Hamiltonian $H^{(r)}$ is given by the natural generalisation; using the commutation relations
\eqref{higherHcomm} for odd $r$ and the already derived result \eqref{NewtonGirard} for even $r$ yields
\begin{align}
E^{(r)} = \left(\pm \epsilon_1\right)^{r} + \left(\pm \epsilon_2\right)^{r} +\dots + \left(\pm \epsilon_S\right)^{r} \ .
\label{Ehigh}
\end{align} 
Since the Hamiltonians commute amongst one another, a single choice of $\pm$ signs fixes all the $E^{(r)}$.

I also show in this section that the relation for the transfer matrix analogous to \eqref{HPsi} is
\begin{align}
\big(u_l+u\big)\, T(u)\, \Psi_{l}  = \,\big(u_l-u\big)\,\Psi_{l}\, T(u) \ .
\label{TPsiT}
\end{align}
Exchanging the order of a raising or lowering operator with the transfer matrix therefore results simply in multiplying by a numerical factor. 
In addition, I show that the transfer matrix can be written in terms of the raising and lowering operators as
\begin{align}
T(u) = \prod_{k=1}^S \left(1-u\epsilon_k \big[\Psi_{k},\,\Psi_{-k}\big]\right)\ .
\label{TPP}
\end{align}
The eigenvalue of the transfer matrix $T(u)$ is then
\begin{align}
\prod_{k=1}^S (1\mp u\epsilon_k)\ .
\end{align}
Here and henceforth I omit the subscript $M$ in the transfer matrix, so $T(u)\equiv T_M(u)$.


Of course, for $\widetilde{H}$ and its corresponding transfer matrix, the same formulas apply with $u_k$ replaced by $\widetilde{u}_k$, the roots of the polynomial $\widetilde{P}_M(\widetilde{u}^2)$ formed from the couplings $\widetilde{b}_m$ in the same fashion as $P_M(u^2)$. Thus the spectrum of $\widetilde{H}$ is
\begin{align}
\widetilde{E}=\pm \frac{1}{\widetilde{u}_1} \pm \frac{1}{\widetilde{u}_1} \pm \dots \pm  \frac{1}{\widetilde{u}_S} \ , 
\label{ffEtilde}
\end{align}
and analogously for the higher Hamiltonians. 

The spectrum of the parity-invariant Hamiltonian $H+\widetilde{H}$ is therefore given by the $2^{2S}=2^{2[(M+2)/3]}$ energies $E+\widetilde{E}$. The dimension of the Hilbert space is $2^{M+2}$ for this $M+2$-site system, and so is exponentially larger than the number of distinct energies.  Thus each energy must be degenerate with an exponentially large multiplicity $2^{M-2[(M-1)/3]}= 2^{S+1}$. The degeneracy is identical for the higher Hamiltonians and hence the transfer matrix.


\subsection{Raising/lowering operators}

The nicest way to find raising/lowering operators with non-zero $\epsilon_k$ is to build them using an ``edge'' operator $\chi_{M+1}$. The edge operator is defined so that it commutes with all the $h_m$ except for the last one: 
\begin{align}
h_M\chi^{}_{M+1} = - \chi^{}_{M+1}h_M\ ,\quad h_m\chi_{M+1} = \chi^{}_{M+1}h_m\quad\hbox{ for }m=1\dots M-1\ .
\label{hchi}
\end{align}
Moreover, $(\chi_m)^2\equiv 1$.
In the explicit representation of the $h_m$ given by \eqref{hh}, one can take e.g.\ $\chi_{M+1}=\sigma^z_{M+2}$, but nothing below depends on this choice. Another choice is given below in order to satisfy a larger algebra (\ref{chialg}) involving $\chi_{M+1}$ and the supersymmetry generators. 
The proofs in this section follow only from the fact properties of the transfer matrix already derived, and do not require any knowledge of the explicit representations of the $h_m$. The relations listed at the beginning of this section are thus valid for {\em all} models whose Hamiltonians can be written in terms of the $h_m$ obeying the algebra (\ref{halg}) extended to include $\chi_{M+1}$ via \eqref{hchi}.

The raising and lowering operators are then simply 
\begin{align}
\Psi_{\pm k} \equiv \frac{1}{N_{k}} T(\mp u_k) \chi^{}_{M+1} T(\pm u_k)\ ,
\label{PsiT}
\end{align}
where the factor $N_k$ is a normalisation specified below. 
Their commutation relations with $H$ follow from an identity derived in the appendix. Namely, 
given the algebra (\ref{hchi}) and the fact that the Hamiltonian commutes with the transfer matrix, 
\begin{align}
[H,T(u)\chi^{}_{M+1}T(-u)] = 2\,T(u)h_{M}\chi^{}_{M+1}T(-u)
\label{HTchiT}
\end{align}
for any $u$. 
The right-hand side obeys the identity
\begin{align}
u\,T(u)h^{}_{M}\chi^{}_{M+1}\,T(-u)= -T(u)\chi^{}_{M+1}\,T(-u) +P_M(u^2)\big(1-uh_{M}\big)\chi^{}_{M+1}\ ,
\label{TchiT}
\end{align}
proven in appendix \ref{app:TchiT} by exploiting the product form (\ref{Gdef}) of the transfer matrix.
The raising and lowering operators now follow simply by exploiting the fact that $P_M(u_k^2)=0$. Setting $u=\mp u_k$ in (\ref{HTchiT},\ref{TchiT}) gives 
\[   \Big[H,T(\mp u_k)\chi^{}_{M+1}T(\pm u_k)\Big]\ =\ \pm \frac{ 2}{u_k} T(\mp u_k)\chi^{}_{M+1}T(\pm u_k)\ ,\]
i.e.\ (\ref{HPsi}) and (\ref{epsilonu}). 
In order to obtain distinct operators, here and henceforth I assume that no $u_k=u_{k'}$ for $k\ne k'$. If two roots do happen to coincide, they can be split by varying one of the $b_m$. The subsequent analysis then applies, and the degenerate case is obtained by taking a limit.

Acting with $\Psi_{\pm k}$ on an eigenstate of $H$ with energy $E_0$ therefore either annihilates the state or gives another eigenstate with energy $E_0\pm \epsilon_k$. 
Since the original Hamiltonian is Hermitian, the $\epsilon_{k}$ and hence the $u_k$ must be real. By convention I take $u_k>0$ for $k=1,2,\dots S$ and often write the raising/lowering operators as $\Psi_l$, where $l=\pm k$ and 
$u_l=\pm u_k$. The definition \eqref{PsiT} requires that
\begin{align}
(\Psi_{l})^2  \propto T(u_l)T(-u_l)= P_M(u_l^2) = 0\ ,
\end{align}
so acting with the same raising or lowering operator twice annihilates any state. These operators thus satisfy Pauli exclusion. Moreover, the $\Psi_{\pm l}$ are the right and left eigenvectors of  $T(u_l)$ with zero eigenvalue:
\begin{align}
T(u_l)\, \Psi_{l} =   \Psi_{-l} T(u_l) = 0 \ .
\label{TPsizero}
\end{align}
Since $T(u)$ is Hermitian, (\ref{PsiT}) also means that $(\Psi_l)^\dagger = \Psi_{-l}$.

The relation (\ref{PsiT}) between the transfer matrix and and the raising and lowering operators makes it straightforward to find the effect of the latter on the eigenvalues of the former. Since the Hamiltonian commutes with $T(u)$ for any $u$, they share the same eigenstates.  Since the $T(u)$ at different values of $u$ commute as well,
\begin{align} 
T(u)\, \Psi_{l}\, T(-u) 
=\frac{1}{N_l} T(-u_l) T(u)\chi^{}_{M+1} T(-u) T(u_l)\ .
\label{TPsiT1}
\end{align}
Using the identity \eqref{TchiT} as well as the definition (\ref{PsiT}) of the raising and lowering operators then gives
\begin{align} 
T(u)\, \Psi_{l}\, T(-u) &=\frac{1}{N_l} T(u_l)\left( -uT(u)h^{}_M\chi^{}_{M+1} T(-u) +P_M(u^2) (1-uh_M)\chi_{M+1}\right)T(-u_l)\cr
&=  -\frac{u}{2} T(u) \big[H,\Psi_l\big] T(-u) + P_M(u^2) \left(\Psi_l - \frac{u}{2} [H,\Psi_l]\right)\cr
&= \frac{1}{u_l}\Big(-u T(u)\Psi_l T(-u) + P_M(u^2) (u_l-u)\Psi_l\Big)\ ,
\end{align}
yielding the elegant formula 
\begin{align}
(u_l+u)T(u)\, \Psi_l\, T(-u)  = P_M(u^2) (u_l- u)\,\Psi_{l}
\label{TPsiT3}
\end{align}
When $u^2$ is not a zero of $P_M(u^2)$, the transfer matrix can be inverted and so \eqref{TPsiT} results.
This formula still holds in the limit $u\to \pm u_k$, where it reduces to \eqref{TPsizero}. 

The commutators of the higher Hamiltonians with the raising/lowering operators are easy to work out using the derivative of the identity \eqref{TPsiT} with respect to $u$, namely
\[(u_l-u)\Psi_{l}T'(u)- (u+u_l)T'(u)\Psi_{l} = \big\{\Psi_{l},\, T(u)\big\}\ . \]
Multiplying by $T(-u)$ on the left and using \eqref{TPsiT} to commute it with $\Psi_l$ gives
\[\big(u^2-u_l^2\big)\Big[T(-u) T'(u)\,,\,\Psi_l\Big] = 2u_l P_{M}(u^2)\Psi_l\ .\]
Expanding $T(-u)T'(u)$ in terms of the higher Hamiltonians using (\ref{higherH}) and comparing the terms at each order in $u$ gives
\begin{align}
\big[H^{(2s-1)},\,\Psi_l \big] = 2(\epsilon_l)^{2s-1} \Psi_l \ ,\qquad\quad \big[H^{(2s)},\,\Psi_l \big] = 0\ .
\label{higherHcomm}
\end{align}
for all positive integers $s$. The $\Psi_l$ are thus raising and lowering operators for all the non-diagonal Hamiltonians.

\subsection{The algebra of the raising and lowering operators}

Here I derive the algebra (\ref{Psialg}), i.e.\ show that the $\Psi_l$ can be viewed as annihilating and creating free fermions. To start, note that
\begin{align}
\big\{\Psi_l,\,\Psi_{-m} \big\}= \frac{1}{N_{m}}\lim_{u\to u_{m}}\big\{\Psi_l, T(u)\chi^{}_{M+1}T(-u) \big\}
\label{antiPsil}
\end{align}
Using (\ref{TPsiT}) to commute the transfer matrix with $\Psi_l$ gives
\[
\big\{\Psi_l,\, T(u)\chi^{}_{M+1}T(-u)\big\} = \frac{u_l+u}{u_l-u}\, T(u) \big\{ \Psi_l,\, \chi^{}_{M+1}\big\} T(-u)
\]
The anticommutator on the right-hand-side can be done by using 
\begin{align}
\big\{ \Psi_l,\, \chi^{}_{M+1}\big\}=
\frac{1}{N_l}\big\{T(-u_l)\chi^{}_{M+1}T(u_l),\, \chi^{}_{M+1}\big\} = \frac{4}{N_l}P_{M-1}(u_l^2)\ ,
\label{etaTP}
\end{align}
which follows by anticommuting $\chi^{}_{M+1}$ with the expression written in the second line of (\ref{etagen}) and the fact that $P_M(u_k^2)=0$. 
Then 
\[\big\{\Psi_l,\, T(u)\chi^{}_{M+1}T(-u)\big\} =  \frac{4}{N_l}P_{M-1}(u_l^2) \frac{u_l+u}{u_l-u} T(u)T(-u)= \frac{4}{N_l}P_{M-1}(u_l^2) P_M(u^2)\frac{u_l+u}{u_l-u}\ .
\]
Taking the limit $u\to u_m$ to get (\ref{antiPsil}) means that this anticommutator vanishes for any $m$ except $m=l$. In this case both the numerator and denominator vanish, and so taking the limit gives
\[ \{\Psi_l,\,\Psi_{-m} \}=-\delta_{lm} \frac{8u_l}{N_{l}^2}P_{M-1}(u_l^2)P'_M(u_l^2)\ .
\]
This gives the algebra (\ref{Psialg}) of the raising and lowering operators when the normalisation is set to  
\begin{align}
(N_l)^{2} = -8 u_l P_{M-1}(u_l^2)  P'_M (u_l^2) = 16 P_{M-1}(u_l^2) \prod_{s=1,s\ne l}^S \left(1-\frac{u_l^2}{u_s^2}\right)\ .
\label{normdef}
\end{align}

\subsection{The Hamiltonian and transfer matrix in terms of raising/lowering operators}

The task of this subsection is to derive (\ref{TPP}) and so give the transfer matrix in terms of the raising/lowering operators. The key to the derivation is showing that, as suggested by their commutation relations \eqref{higherHcomm}, all the Hamiltonians can be written nicely in terms of the bilinears as
\begin{align}
H^{(2t+1)} = \sum_{k=1}^S \epsilon_k^{2t+1} \big[ \Psi_{k},\,\Psi_{-k} \big]
\label{Hee}
\end{align}
for $t$ any non-negative integer.

The contour-integral definition of the $H^{(t)}$ gives an expression for the Hamiltonians useful in this derivation. Since ${\cal H}(u)$ is meromorphic, this contour integral can be rewritten by changing variables to  $\epsilon=1/u$, where the contour in $\epsilon$ is now at a circle at some very large $|\epsilon|$, encircling all the poles at  $\epsilon=\pm\epsilon_k$. Thus
\[H^{(r)} =\frac{1}{2\pi i}\oint d\epsilon\, \epsilon^{r-2} {\cal H}(u)= -\frac{1}{2\pi i}\oint d\epsilon \frac{\epsilon^{2S+r-2}}
{\prod_{k=1}^{S}(\epsilon^2-\epsilon_k^2)}T(-1/\epsilon)T'(1/\epsilon)\ .\]
Since the maximum power of $u$ in $T(u)$ is $S$, for positive $r$ the integrand has no pole at $\epsilon=0$, only those at $\pm\epsilon_k$. Thus
\begin{align}
H^{(r)} &=- \frac{1}{2}\sum_{k=1}^S  \frac{\epsilon_k^{2S+r-3}}
{\prod_{s=1,s\ne k}^{S}(\epsilon_k^2-\epsilon_s^2)}\left(T(-1/\epsilon_k)T'(1/\epsilon_k) - (-1)^r T(1/\epsilon_k)T'(-1/\epsilon_k) \right)\cr
&= \sum_{k=1}^S \frac{u_k^{-r}}{P'_M(u_k^2)}\left(T(-u_k)T'(u_k) - (-1)^r T(u_k)T'(-u_k) \right)
\label{HTT}
\end{align}
For $r$ even, using the identity (\ref{Pprime}) shows that (\ref{HTT}) indeed reduces to \eqref{NewtonGirard}. 

Thus to prove \eqref{Hee}, consider
\begin{align}
\big[ \Psi_k,\,\Psi_{-k} \big]= \frac{1}{N_{k}}\lim_{u\to u_{k}}\big[\Psi_k, T(u)\chi^{}_{M+1}T(-u) \big]
\label{commPsil}
\end{align}
Using 
(\ref{TPsiT}) to commute the transfer matrix with $\Psi_l$ gives
\[
\big[\Psi_k,\, T(u)\chi^{}_{M+1}T(-u)\big] = \frac{u_k+u}{u_k-u}\, T(u) \big[ \Psi_k,\, \chi^{}_{M+1}\big] T(-u)
\]
Since this denominator is vanishing as $u\to u_k$, taking this limit in (\ref{commPsil}) requires some care. The zero eigenvectors in (\ref{TPsizero}) come to the rescue, guaranteeing the numerator also vanishes in this limit, giving 
\begin{align*}\big[ \Psi_k,\,\Psi_{-k} \big] &=- 2\frac{u_k}{N_k} 
\left(T'(u_k)\Psi_k\chi^{}_{M+1}T(-u_k) + T(u_k)\chi^{}_{M+1}\Psi_k T'(-u_k)\right)\ .
\end{align*}
The order of $\chi^{}_{M+1}$ and $\Psi_l$ in each of the two terms can be switched by using
the anticommutator (\ref{etaTP}). The terms involving $\Psi_k$ then vanish because of the zero modes in (\ref{TPsizero}) so that
\begin{align}
\big[ \Psi_k,\,\Psi_{-k} \big]
=\frac{1}{P'_{M}(u_k^2)}
\Big( T'(-u_k) T(u_k) + T(-u_k) T'(u_k)\Big)
\label{PsiPsicomm}
\end{align}
using the expression \eqref{normdef} for the normalisation.
Using (\ref{PsiPsicomm}) in (\ref{HTT}) then gives instantly \eqref{Hee}, and as a special case, the expression (\ref{He}) of the Hamiltonian in terms of the bilinears.


Relating the transfer matrix to the bilinears $\Psi_l\Psi_{-l}$ as in (\ref{TPP}) is now straightforward to do. The algebra (\ref{Psialg}) requires that all these bilinears commute with each other and that each squares to itself. Moreover,
\[ \left(\big[\Psi_k,\,\Psi_{-k}\big]\right)^2 = \big\{\Psi_k,\,\Psi_{-k}\big\} = 1\ .\]
The expressions (\ref{NewtonGirard}) and (\ref{Hee}) then can be unified to give 
\begin{align}
H^{(r)} = \sum_{k=1}^S \left(\epsilon_k\,\big[ \Psi_{k},\,\Psi_{-k} \big]\right)^r\ 
\label{Hee2}
\end{align}
for all positive integers $r$.
The sum of the higher Hamiltonians is then related to the transfer matrix via their definition (\ref{higherH}) as
\[-\frac{d}{du} \ln T(u)= \sum_{r=1}^{\infty} H^{(r)}u^{r-1} =\sum_{k=1}^S \frac{\epsilon_k \big[ \Psi_{k},\,\Psi_{-k} \big]}
{1-u\epsilon_k \big[ \Psi_{k},\,\Psi_{-k} \big]}=-\frac{d}{du}\ln\prod_{k=1}^S\left(1-u\epsilon_k \big[ \Psi_{k},\,\Psi_{-k} \big]\right)
\ .\]
This differential equation in $u$ is valid in the region near $u=0$. Solving it using $T(0)=1$ gives at last the elegant expression \eqref{TPP}.  As a consistency check, note that the commutation relation (\ref{TPsiT}) follows simply from the facts that 
\[ [\Psi_k,\,\Psi_{-k}]\Psi_{\pm k}=\pm \Psi_{\pm k}\ ,\qquad \Psi_{\pm k}[\Psi_k,\,\Psi_{-k}]=\mp \Psi_{\pm k}, \]
as follows from the algebra \eqref{Psialg}.

\section{Supersymmetry and its extension}
\label{sec:susy}

The exponentially large degeneracies described in section \ref{sec:spectrum} come as rather a surprise. They occur in this model only for open boundary conditions; it is easy to check using exact diagonalization that taking periodic boundary conditions splits the degeneracies without spoiling the integrability. Whereas such degeneracies do occur in integrable models, having them identical for each level is rather unusual. I show in this section that this behaviour is a consequence of a symmetry algebra extending the supersymmetry algebra. Its generators commute with the transfer matrix but not each other, and not with the raising/lowering operators. 


\subsection{Supersymmetry}

Although not obvious via its formulation in terms of spins, any Hamiltonian whose generators obey (\ref{halg}) in general has a supersymmetry analogous to that appearing in particle physics, where the Hamiltonian itself is part of the supersymmetry algebra. Precisely, it is supersymmetric whenever its generators can be written in terms of fermionic operators $\chi_m$  obeying the algebra
\begin{align}
\chi_m^2 = a_m^2 ;\qquad  \chi^{}_m \chi^{}_{m+1}=\chi^{}_{m+1}\chi^{}_m\ ;\qquad \chi^{}_m \chi^{}_n = - \chi^{}_n \chi^{}_m\ \hbox{ for }\ |n-m|>1\ ,
\label{chialg}
\end{align}
where $a_m$ is a real number. Setting
\begin{align}
h_{m}\equiv\chi_{m-1}\chi_m
\end{align}
means that (\ref{chialg}) immediately implies that the $h_m$ obey (\ref{halg}) with $b_m^2=(a_{m-1}a_m)^2$.
For open boundary conditions, the index on $\chi_m$ runs over $m=0,\cdots, M+1$, although the Hamiltonian does not involve $\chi_{M+1}$; this ``extra'' operator is involved only in the construction of the raising and lowering operators in section \ref{sec:open}.  For simplicity, $a_{M+1}\equiv 1$. 
The Hamiltonian for open boundary conditions is therefore
\begin{align}
H= \sum_{m=1}^{M} \chi^{}_{m-1}\chi^{}_{m}=\sum_{m=1}^{M} h_m\ .
\label{Hdef}
\end{align}

The basic fermionic conserved charges are constructed in a similar fashion to those found in various chains in \cite{Fendley03a}. Here, there are two such ``supercharges'', consisting of sums of the $\chi_a$ with even and odd indices:
\begin{align}
O=\sum_{j=0}^{[(M+1)/2]} \chi^{}_{2j+1}\ ;\qquad\quad
E=\sum_{j=0}^{[M/2]} \chi^{}_{2j}\ .
\label{QoQe}
\end{align}
Since each term in $O$ anticommutes with each of the others, its square is proportional to the identity, and likewise for $E$:
\begin{align}
(O)^2= \sum_{j=1}^{[(M+1)/2]} a_{2j-1}^2\ ; \qquad\quad 
(E)^2= \sum_{j=1}^{[M/2]} a_{2j}^2\ .
\label{AsqBsq}
\end{align}
Defining the Hamiltonian in the usual supersymmetric fashion \cite{Witten82} gives the generalisation of \eqref{HQ} to non-uniform couplings:
\begin{align}
H=\frac{1}{2}(E+O)^2 - \frac{1}{2}\sum_{m=0}^{M} a_m^2= \frac{1}{2} 
\{E,O\}\ .
\label{HQQ}
\end{align}
Since $E^2$ and $O^2$ are constants, (\ref{HQQ}) requires that 
[O,H]=[E,H]=0.
This supersymmetry is also present for appropriately twisted boundary conditions when the generators are interpreted mod $M$. However, I remain focused on open boundary conditions here.

It is easy to find explicit $\chi_m$ for each of the Hamiltonians in (\ref{hh}). For a collection of $L$ spins, $2L$ Majorana fermion operators are defined in the usual fashion in \eqref{JW}. 
Two sets of $\chi_m$ then are
\begin{align}
\chi_m =ia_m\,\psi_{2m+2}\psi_{2m+3}\psi_{2m+4}\ ;\qquad\quad 
\widetilde{\chi}_m = i\widetilde{a}_{m}\, \psi_{2m+1}\psi_{2m+2}\psi_{2m+3}\ ,
\label{chi4fermi}
\end{align}
for $m=0\dots M$ and $M=L-2$. The operator used in constructing the raising and lowering operators can be taken to be 
$\chi_{M+1}=\psi_{2m+4}$. 
The $\chi_m$ obey (\ref{chialg}) among themselves as do the $\widetilde{\chi}_m$, while the different types anticommute:
\[ \{ \chi_m,\,\widetilde{\chi}_{m'}\} = 0\qquad\hbox{ for all }m,m'\ .\]
Defining $\widetilde{E}$ and $\widetilde{O}$ analogously to $E$ and $O$ gives two commuting supersymmetric Hamiltonians defined using (\ref{HQQ}), namely those given in (\ref{hh}) and (\ref{HH}). Written in terms of the fermions, each is purely four-fermion:
\begin{align}
H= \sum_{m=1}^M b_m\,\psi^{}_{2m}\psi^{}_{2m+1}\psi^{}_{2m+3}\psi^{}_{2m+4}\ ,\quad\qquad
\widetilde{H}= \sum_{m=1}^M \widetilde{b}_m\,\psi^{}_{2m-1}\psi^{}_{2m}\psi^{}_{2m+2}\psi^{}_{2m+3}\ ,
\label{Hferm}
\end{align}
where $b_m=-a_{m-1}a_m$ and $\widetilde{b}_m=-\widetilde{a}_{m-1}\widetilde{a}_m$.

An obvious symmetry of the Hamiltonians $H$ and $\widetilde{H}$ is ``fermion-number parity'' or equivalently, flipping all the spins. The generator ${\cal F} $ obeys ${\cal F}^2=1$ and can be written in terms of the spins and fermions as
\begin{align}
{\cal F}\equiv \prod_{l=1}^L\sigma^x_l = (-i)^L\prod_{j=1}^{2L}\psi^{}_l\ .
\end{align}
The spectrum can then be grouped into sectors with eigenvalue $\pm 1$ of ${\cal F}$. 
Since the supercharges each anticommute with ${\cal F}$ and square to a non-zero constant, they map between the two sectors. Since they also commute with $H$, the spectrum in each of these sectors must be identical. In other words, each of the supercharges is a strong zero mode.

The supercharges commute with all the symmetry generators $\QQ{s}$ and hence the full transfer matrix, just as the Hamiltonian does. The proof is similar. A key identity is
\begin{align}
\chi_m\chi_{m+1}\chi_{m+2}=\chi_{m}h_{m+2}=-h_{m+2}\chi_{m}=-\chi_{m+2}h_{m+1} = h_{m+1}\chi_{m+2}\ .
\label{hchiident}
\end{align}
while otherwise
\begin{align}
\big[\chi_m,\,h_{m'}\big]=0\qquad\hbox{ for }m\ne m'+2,\,m'-1\ .
\end{align}
Then 
\[
\big[\chi_{m}, h_{m_1}\,h_{m_2}\,\dots\,h_{m_s}\big]=
\begin{cases}
 2\chi_{m} h_{m_1}\,h_{m_2}\,\dots\,h_{m_s} &\quad 
m-m_r=1,-2,\  a\ne m_{r+1}-1, \ a\ne m_{r-1}+2 \\
\ 0&\quad\hbox{otherwise }.
\end{cases}
\]
The additional restrictions arise because $[\chi_m,\,h_{m-1}h_{m+2}]=0$. Doing the sums over $m$ and the $m_r$, the terms cancel pairwise because
\[ \big[\chi_m,\,h_{m+1}\big] + \big[\chi_{m+2},\,h_{m}\big] = 0\ .\]
as a consequence of \eqref{hchiident}. 
All terms in each of $[O,\,\QQ{s}]$ and $[E,\,\QQ{s}]$ pair in this fashion because of the additional restrictions, and because each pair involves $\chi_m$ and $\chi_{m+2}$. Thus both $E$ and $O$ are conserved charges:
\begin{align}
\big[ E,\,T(u)\big]=\big[ O,\,T(u)\big]= 0\ .
\end{align}


\subsection{An equivalent Hamiltonian}
\label{sec:newham}

The solution of the open chain in section \ref{sec:open} using raising and lowering operators requires only a set of generators $h_m$ obeying the algebra (\ref{halg}). Since having fermionic operators obey (\ref{chialg}) implies both (\ref{halg}) and the supersymmetry, it is straightforward to find other models with the same physics. Here I present such a model where $H$ stands alone; i.e.\ there is no analogous commuting Hamiltonian $\widetilde{H}$. 

This equivalent Hamiltonian is easiest to express in terms a Majorana chain coupled to an Ising chain. The Majorana operators $\psi_j$ are defined as above on a chain of $L$ sites, while the Ising chain is comprised of $2L$ sites, each having a two-state system. Denoting the Pauli operators acting non-trivially on the latter chain by $\tau^a_j$, the $\chi_m$ defined by
\begin{align}
\chi^{}_{3j-3}= a_{3j-3}\,\psi_{j}\tau^x_{j}\ ,\qquad
\chi^{}_{3j-2}=a_{3j-2}\,\psi^{}_{j}\ ,\qquad
\chi^{}_{3j-1}=a_{3j-1}\,\psi^{}_{j}\tau^{z}_{j}\tau^{z}_{j+1}
\label{chinew}
\end{align}
satisfy the algebra (\ref{chialg}) for $m=0\dots 6L-2$. Thus I take $M=6L-2$ with the edge operator defined as $\chi_{M+1}\equiv \psi^{}_L\tau^z_L$. The supersymmetry generators and Hamiltonian are defined via (\ref{QoQe}) and (\ref{HQQ}) as above, giving
\begin{align}
H_{\rm two-chain} = \sum_{j=1}^{2L} b_{3j-2}\tau^x_j 
+ \sum_{j=1}^{2L-1}\left(b_{3j-1} \tau^z_j\tau^z_{j+1} + ib_{3j}\psi^{}_j\psi^{}_{j+1}\tau^{z}_{j}\tau^{y}_{j+1}\right)
\label{H2}
\end{align}
Rewriting the Majorana chains in terms of spins means the bilinear $i\psi_j\psi_{j+1}$ corresponds to $\sigma^x_j$ for $j$ odd, and $\sigma^z_j\sigma^z_{j+1}$ for $j$ even.
When $b_{3j}=0$, the two chains decouple and the Hamiltonian is precisely that of the transverse-field Ising model, and independent of the Ising chain. In this limit, the decoupled chain thus gives a simple explanation for the identical degeneracies at every energy.

The Hamiltonian \eqref{H2} has only one set of raising and lowering operators, because there is no second set of commuting generators. The complete spectrum thus is given by (\ref{EEu}), where the $u$ remain zeroes of the same polynomial $P_M(u)$ defined by the $b_m$. Moreover, the degeneracies for open boundary conditions are essentially the same. In \eqref{chinew}, there are now $6L$ different $\chi_m$ for a system with $3L$ sites, i.e.\ twice as many supersymmetry generators as there are sites, just as in the four-fermi case with Hamiltonian $H+\widetilde{H}$. There are $2^{[(M+2)/3]}=2^{2L}$ distinct energies with a Hilbert space of dimension $2^{3L}$. The degeneracy per level is therefore $2^{L}$ even for non-vanishing $b_{3j}$.

In some ways this coupled-chain system resembles that introduced in \cite{Grover14} to provide a lattice model with emergent supersymmetry. Both feature an Ising chain coupled to a Majorana chain, where the number of Majorana fermion operators in the latter is the same as the number of spins in the former. Both feature a self-duality for uniform couplings, and as discussed below in section \ref{sec:staggered}, a multicritical point with a flow to Ising with supersymmetry spontaneously broken. However, the couplings between the two chains  are different, leading here to the multicritical point $z=3/2$ point analysed below, whereas that of \cite{Grover14} leads to the tricritical Ising point, described by a $z=1$ (but not free-fermionic) conformal field theory in the continuum limit.


\subsection{Extended supersymmetry}

The operators $E$ and $O$ commute with the Hamiltonian and transfer matrix, but not with each other. Their anticommutator gives $H$, while their commutator thus gives a bosonic conserved charge comprised of non-local operators.  An important but less obvious feature is that the latter can be split into two pieces, each of which commutes with $T(u)$. I show here how this feature can be used to construct a hierarchy of bosonic and fermion operators obeying a generalisation of the supersymmetry algebra.

The bosonic conserved charge $[E,O]$ can be split as
\begin{align}
\frac{1}{2}\big[E,\,O\big] &= \A{2} - \B{2}\ ,\cr
\A{2}\equiv  \sum_{j>k+1}
\chi^{}_{2k}\chi^{}_{2j-1}\ , \quad &\quad
\B{2} \equiv  \sum_{j>k} \chi^{}_{2k-1}\chi^{}_{2j}\ .
\label{A2def}
\end{align}
where the upper limit of any sum is chosen to make all $\chi_m$ have $m\le M$.  It is easy to check explicitly using \eqref{hchiident} that $H$ commutes with $\A{2}$ and $\B{2}$ individually. More generally,
\[ \big[[E,O],\,T(u)\big] =0= \Big[\A{2},T(u)\Big] + \Big[\B{2},T(u)\Big]\ .\]
There can be no cancellation between the latter two expressions because
\[  \chi^{}_{2k}\chi^{}_{2j-1} h_{m_1}\dots h_{m_s} \ \ne\ \pm   \chi^{}_{2k'-1}\chi^{}_{2j'} h_{m'_1}\dots h_{m'_s}\]
for any choice of $j>k+1$, $j'>k'$, and the $m_r$ and $m'_{r'}$.  Therefore
\begin{align}
\Big[\A{2},T(u)\Big] = \Big[\B{2},T(u)\Big]=0\ .
\label{ABT0}
\end{align}

These bosonic conserved quantities $\A2$ and $\B2$ neither commute with each other nor with $Q$. Thus a huge zoo of quantities commuting with $H$ can be constructed by taking products of $O$, $E$, $\A2$ and $\B2$, and then splitting into the even and odd parts as in \eqref{A2def}.  A useful hierarchy of such operators $\A{s}$ and $\B{s}$ is similar to $\A{2}$ and $\B2$, in that the indices on the $\chi_m$ in each term in the sum must differ by at least three, and alternate between even and odd. The label $s$ simply is the number of $\chi_m$ present, so that having $s$ odd gives a fermionic charge, while $s$ even bosonic. The easiest way to prove that such operators are indeed charger is to express them as commutators or anticommutators. For example, the next charges are given by
\begin{align}
\A{3} &= \frac{1}{2} \big\{O,\A{2} \big\} = \frac{1}{2} \big\{ O,\B{2} \big\} = \sum_{j>k+1>j'+1} \chi^{}_{2j'-1}\chi^{}_{2k} \chi^{}_{2j-1}\ ,\cr
\B{3} &= \frac{1}{2} \big\{E,\B{2}\big\} = \frac{1}{2} \big\{E,\A{2} \big\} = \sum_{j>k>j'+1} \chi^{}_{2j'}\chi^{}_{2k-1} \chi^{}_{2j}\ .
\end{align}
Each therefore must commute with the Hamiltonian and transfer matrix.

The trick for even $s$ in general is to split a commutator or anticommutator into the difference $\A{s}-\B{s}$, where $\A{s}$ contains all the terms where the highest index $m$ on a $\chi^{}_m$ in that term is odd, while $\B{s}$ contains all highest indices even. Thus for example
\[ \big[E,\A{3}\big] = -\big[O,\B{3}\big] = 2\A{4}-2\B{4}\ , \]
and it is easy to see that all terms obey the rule that all indices differ by at least three and alternate between even and odd. Both $\A{4}$ and $\B{4}$ commute with $T(u)$ by the same argument leading to \eqref{ABT0}. 
Continuing in this fashion gives the odd ones as anticommutators. Letting $\A0=\B0=1$,  $\A1=E$ and $\B1=O$ gives
\begin{align}
\A{2r+1} \equiv \frac{1}{2} \big\{O,\A{2r} \big\} = \frac{1}{2} \big\{ O,\B{2r} \big\}\ ,\qquad
\B{2r+1} \equiv \frac{1}{2} \big\{E,\A{2r} \big\} = \frac{1}{2} \big\{ E,\B{2r} \big\}\ \ .
\end{align}
The even ones need to be split: 
\begin{align} 
\A{2r}-\B{2r} \equiv \frac{1}{2} \big[E,\A{2r-1}\big]=  -\frac{1}{2} \big[O,\B{2r-1}\big]\ .
\label{QQdef}
\end{align}
All terms in these currents obey the property that the indices are at least three apart. These defining relations use the ``unnatural'' commutator or anticommutator; for example the bosonic charge $\A{2}$ is obtained from the {\em commutator} of two fermionic charges $O$ and $E$. The remaining unnatural ones are simply 
\begin{align}
\big[O,\A{2j+1}\big]=0\ ,\qquad\quad
\big[E,\B{2j+1}\big]=0\ \ ,
\label{QQdef2}
\end{align}
because $O^2$ and $E^2$ are constants. 

The charges $\A{s}$ and $\B{s}$ do not commute or anticommute among themselves like the $\QQ{s}$ do.  A few of the relations are simple; for example, \[  \frac{1}{2} \big\{\A{2},\B{2}\big\} = \QQ{2} + \A{4} + \B{4}\ . \]
In general, however, the commutators/anticommutators are not so nice, and so still more conserved charges are generated. I have not found a closed algebra, nice or not, like that that found in \cite{Fendley06} for the Cooper pair model.


\comment{
The conserved charges defined by (\ref{QQdef}) can also be defined in terms of recursion relations. 
To do this, I define the truncated operators $\QQ{r}_{2k-1}$ to be $\A{r}$ restricted to involve only $\chi^{}_m$ with $m\le 2k-1$. Likewise, the $\QQ{r}_{2k}$ are the $\B{r}$
restricted to involve only only $\chi^{}_m$ with $m\le 2k$. For example, the supersymmetry generators are truncated to
\begin{align}
\QQ{1}_{2k-1}= \sum_{j=1}^{k} \chi^{}_{2j-1}\ ,\qquad\quad \QQ{1}_{2k}= \sum_{j=1}^{k} \chi^{}_{2j}
\end{align}
Thus the supercharges for the full system are $O=\QQ{1}_{M-1}$ and $E=\QQ{1}_{M}$ for $M$ even and vice-versa for $M$ odd.
The recursion relations for any $M$, $r$ are then
\begin{align}
\QQ{r}_M= \QQ{r}_{M-2}+ \chi^{}_{M}\QQ{r-1}_{M-3}  
\label{Qrecur}
\end{align}
Since the indices of the $\chi^{}_a$ in each term are necessarily at least three apart, there is a maximum possible value of $r$ for a given $m$ in $\QQ{r}_m$, e.g.\ for $m=7$, $r_{\rm max}=3$. If $\QQ{r}_m$ with $r>r_{\max}$  appears, it is defined to be zero, and likewise for $r<0$, $m<0$ or $m>M$. }


\subsection{Degeneracies from the extended supersymmetry algebra}

The charges  $\A{s}$ and $\B{s}$ commute with the Hamiltonian and transfer matrix, but in general not the raising and lowering operators. They thus can be used to construct other raising and lowering operators defined by
\begin{align}
{\Psi}^{(2r)}_l\equiv \frac{1}{2}\Big\{\A{2r-1}+\B{2r-1},\,\Psi_l\Big\}\ ,\qquad\quad
{\Psi}^{(2r+1)}_l\equiv \frac{1}{2}\Big[\A{2r}+\B{2r},\,\Psi_l\Big]\ .
\label{Psisdef}
\end{align}
Defining  $\A{0}=\B{0}=1$ gives ${\Psi}^{(1)}_l=\Psi_l$. Taking the appropriate commutators and anticommutators with \eqref{HPsi} and \eqref{TPsiT} then gives
\begin{align}
\Big[H,\,{\Psi}^{(s)}_l\Big] = 2\epsilon_l {\Psi}^{(s)}_l\ ,\qquad\quad
\big(u_l+u\big)\, T(u)\,{\Psi}^{(s)}_l = \,\big(u_l-u\big)\,{\Psi}^{(s)}_l\, T(u) 
\label{TPsiT2}
\end{align}
for all $s$.

The $\widetilde{\Psi}^{(s)}_l$ do not satisfy the full free-fermion algebra in general, but their satisfying \eqref{TPsiT2} is sufficient to make them raising and lowering operators, either annihilating a eigenstate of $H$ or mapping it to another with energy shifted by $\epsilon_l$.  These operators are non-trivial and distinct for $s\le S$, as is easy to check using the explicit expressions of the charges in terms of the $\chi_m$ and the fact that $\chi_{M}$ commutes with $\chi_{M+1}$. Thus for at least some energy eigenstates $|E\rangle$ and $s\ne s'$,
\begin{align}
\Psi^{(s)}_l |E\rangle\ \ne\ \Psi^{(s')}_l|E\rangle\ ,
\label{PsiE}
\end{align}
because energy eigenstates form a complete set of states.  The energy shift however is independent of $s$, so the distinct states in \eqref{PsiE} must be degenerate. The states therefore form degenerate multiplets, as typical in the presence of any non-abelian symmetry algebra. Since for a given $l$ all $S$ operators $\Psi^{(s)}_l$ are distinct and, it is natural to expect that the dimensions of these multiplets are exponentially large in $S$ and hence the system size. Such degeneracies indeed occur in the spectra of $H$ and $T(u)$, as described in section \eqref{sec:spectrum}, and thus are a natural consequence of the symmetry generators $\A{s}$ and $\B{s}$. 

Characterising the multiplets precisely from representation-theory point of view seems a slightly ambitious task, given that that algebra involving all the $\A{s}$ and $\B{s}$ is not known.  However, a remarkable result of the earlier analysis is that this representation theory seems to be very simple, in that the size of the multiplets is completely independent of state, as showing in section \eqref{sec:spectrum}. This strongly suggests that even though the $\Psi^{(s)}$ defined in \eqref{Psisdef} do not have nice commutation/anticommutation relations, there exists some subalgebra of the full algebra which does. Moreover, given the structure of degeneracies, it is natural to expect that it is some sort of Clifford algebra.

\section{Staggered and uniform couplings}
\label{sec:staggered}

The analysis above has been rather formal, and applies to any couplings $b_m$ as long as the generators of the Hamiltonian obey \eqref{halg}.  The upshot is the complete spectrum is determined simply in terms of the roots $u_k^2$ of the polynomial $P_M(u^2)$ of order $u^{2[M/3]}$. Here I describe some of the physical properties of the spectrum when the couplings are staggered. The easiest form of staggering to handle is where they repeat every third term in the Hamiltonian, i.e.\ 
\begin{align}
b_{3j+1}=b_{1}\equiv\sqrt{\alpha},\qquad b_{3j+2}=b_{2}\equiv\sqrt{\beta},\qquad  b_{3j+3}=b_{3}\equiv\sqrt{\gamma}, \ .
\label{stagg}
\end{align}
This form of staggering is quite natural from the point of view of the commutation relations, since each all elements of the set $\{h_{j},\,h_{j+3},\,h_{j+6},\dots \}$ commute with one another. Tt also is quite natural in the Hamiltonian $H_{\rm two-chain}$ in \eqref{H2}, where $\gamma$ provides a way of tuning the coupling between the two chains, with $\alpha$ and $\beta$ the Ising couplings usually denoted $h$ and $J$.

Finding the dispersion relation for the excitations amounts to finding the roots of $P_M(u^2)$ in the special case \eqref{stagg}. I give this derivation in appendix \ref{app:disper}, parametrising the roots in terms of two variables, $B$ and $p$.  Because the interactions are next-nearest neighbour, $B$ ends up being related to $p$ by the cubic equation
\begin{align}
B^3 =2\alpha\beta\gamma \cos p + (\alpha\beta+\alpha\gamma+\beta\gamma)B\ .
\label{Bp1}
\end{align}
The solution needed is the one where $B$ is real and obeys
\begin{align}
B> (\alpha\beta\gamma)^{1/3} \ .
\label{Bge}
\end{align}
The resulting explicit expression for $B(p)$ in general is not terribly illuminating. 
The variable $0\le p\le \pi$ is akin to the momentum for this open-chain problem, and for $S$ large satisfies the standing-wave condition 
\begin{align}
e^{2ipS} = \frac{B^3e^{-ip}+\alpha\beta\gamma}{B^3e^{ip}+\alpha\beta\gamma}
\label{pquant}
\end{align}
where $M=3S$ is a multiple of three. When $M$ is not a multiple of $3$, the right-hand side of \eqref{pquant} is modified, but remains a phase. Given $B$ and $p$, the expression for the energy levels is  
\begin{align}
\epsilon^2 = \frac{1}{\alpha\beta\gamma B^2}\big(B^2-\alpha\beta\big)\big(B^2-\alpha\gamma\big)\big(B^2-\beta\gamma\big)\ .
\label{disper}
\end{align}

The dispersion relation $\epsilon(p)$ thus can be found simply by solving the cubic equation \eqref{Bp} to get $B(p)$ and substituting into (\ref{disper}). The standing-wave condition \eqref{pquant} determines the precise values of $p$ allowed, which are labelled by the integers $k=1,\dots, S$ in the foregoing. For large $S$, the splitting between successive solutions of \eqref{pquant} is $\Delta p\sim \pi/S$, and so the density of states is simply $3\pi/M$ to leading order in $1/M$. The ground-state energy per site is therefore 
\begin{align}
\frac{E_0}{M} \approx -\frac{1}{3\pi}\int_0^\pi dp\, \epsilon(p)\ .
\end{align}
Excitation energies simply correspond to changing any of the signs in \eqref{ffE} from negative to positive, and thus increasing the energy by $2\epsilon_k$ when the $k$th sign is chosen. 

\begin{figure}
\begin{center}
\includegraphics[scale=0.38]{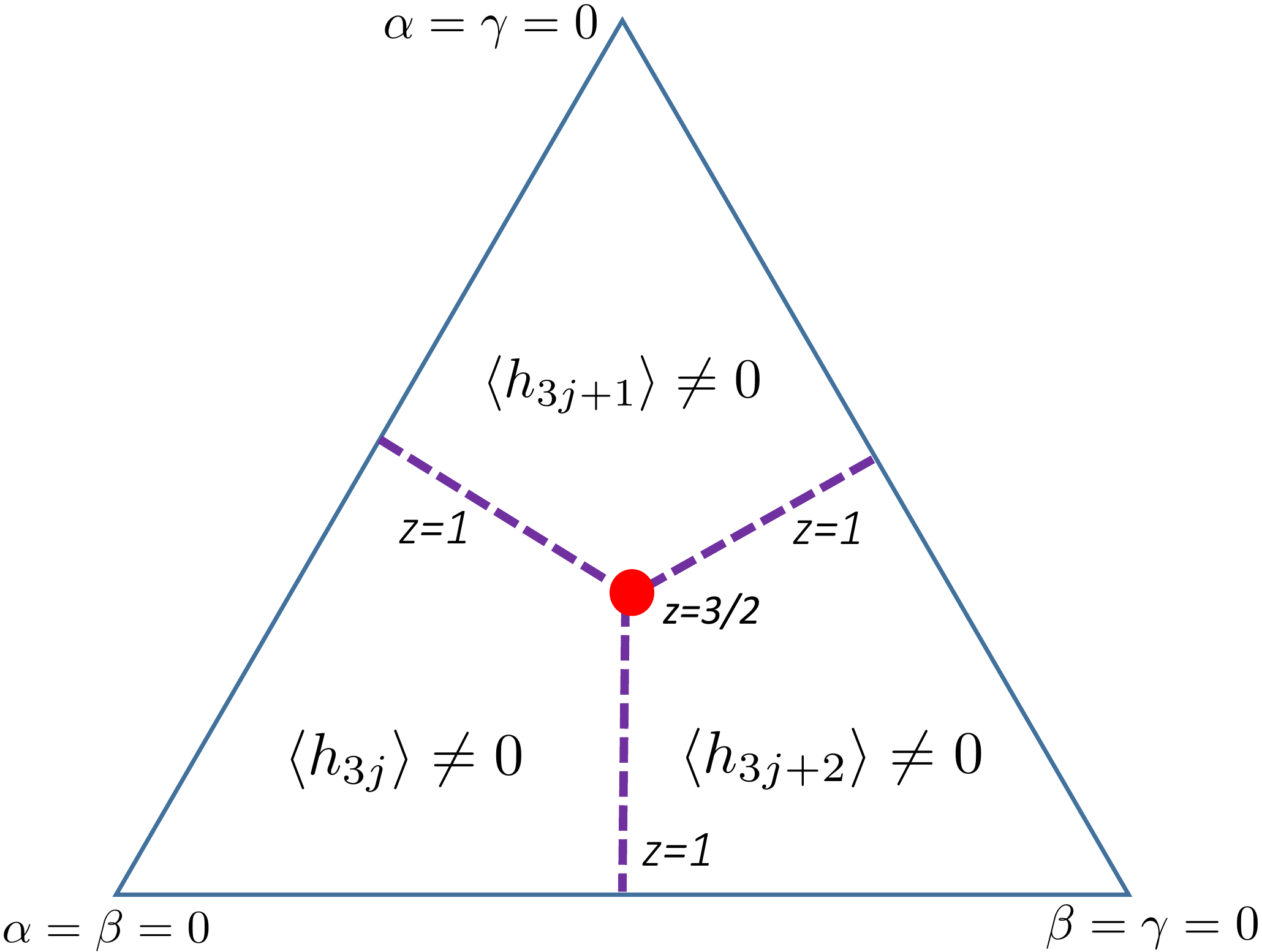}
\end{center}
\caption{\small The phase diagram with staggered couplings, with $\alpha+\beta+\gamma=3$. Each side of the triangle has one vanishing coupling, and each corner two. The multicritical point at $\alpha=\beta=\gamma=1$ has dynamical critical exponent $z=3/2$, while the dashed lines are Ising-type transitions with $z=1$ that separate gapped phases.}
\label{fig:phasediagram}
\end{figure}

Gapless excitations arise when $\epsilon(p)\to0$ in some limit.  In appendix \ref{app:disper} they are shown to occur only when $p\to\pi$, and moreover, only if two of the couplings $\alpha,\beta,\gamma$ are equal. Setting $\alpha=\beta=1$, the cubic equation \eqref{Bp} for $B_\gamma\equiv B(\pi)$ reduces to
\[ \big(B_\gamma-1\big)\left(B_\gamma^2+B_\gamma-2\gamma\right)=0\ .\]
Only for $\gamma\le 1$ does the solution $B_\gamma=1$ obey the inequality \eqref{Bge}, so 
\begin{align}
B_\gamma= 
\begin{cases}
\frac{1}{2}\big(\sqrt{8\gamma+1}-1\big)\ ,\qquad&\gamma\ge 1\cr
1\ ,&\gamma\le 1\ .
\end{cases}
\label{Bgamma}
\end{align}
Utilising \eqref{disper} means that $\epsilon(p)\to 0$ as $p\to\pi$ only for $\gamma\le \alpha=\beta$. It is shown in the appendix that $\epsilon(p)$ vanishes linearly here. Thus these line segments correspond to having a critical phase transition. For $\gamma>\alpha=\beta$, the system is gapped.

The phase diagram of the staggered model in therefore given in figure \ref{fig:phasediagram}, where for convenience the couplings are normalized to satisfy $\alpha+\beta+\gamma=3$. At the corners of the triangle, two couplings vanish, and the Hamilton is trivial to diagonalize. The order parameters for the three phases are therefore simply expectation values of $h_m$, as shown in the figure. In the original four-fermion model, this amounts to favouring alignment among spins on two of the three sublattices, e.g.\ n the sites $m=3j$ and $3j+2$ in the phase at the top of the diagram. The critical line segments are therefore typical order-disorder transitions; symmetry between sublattices (broken only by being on an open chain) implies they happen when two couplings are equal. 

More useful intuition comes from considering the coupled-chain Hamiltonian \eqref{H2}. Setting $\gamma=0$ there corresponds to decoupling the two chains, leaving an interacting Ising chain and a Majorana chain with trivial Hamiltonian.  Thus along the bottom of the triangle, the physics is simply that of the Ising chain, which indeed has a critical transition when the two couplings are equal. The symmetry under exchange $\alpha\leftrightarrow\beta$ is simply Kramers-Wannier duality. Turning on $\gamma$ is a (marginally) irrelevant perturbation, leaving the transition intact for $\gamma<\alpha=\beta$.  

The Hamiltonian $H_{\rm u}$ with uniform couplings is therefore a multicritical point where the three critical line segments meet. The physics here turns out to be rather striking. 
Here the solution to \eqref{Bp} is simply
\begin{align}
B_{\rm u} =  2\cos\frac{p}{3}\ ,
\label{Buniform}
\end{align}
with $0\le p <\pi$. The dispersion relation \eqref{disper} in the uniform case is then exactly
\begin{align}
\epsilon^2_{\rm u}(p) =\frac{\sin^3 p}{\sin\frac{p}{3}\sin^2\frac{2p}{3}}\,
\end{align}
so indeed $\epsilon(p)\to 0$ as $p\to\pi$. However, the dispersion relation at the fermi sea is not linear. Instead, for $p$ near $\pi$,
\begin{align}
\epsilon_{\rm u}(p)\approx \left(\frac{4}{3}\right)^{3/4} |\pi-p|^{3/2}\ .
\label{gaplessdisper}
\end{align}
Since $\Delta p \propto 1/L$ for large system size $L$, the gapless dispersion relation (\ref{gaplessdisper}) corresponds to having dynamical critical exponent $z=3/2$. The critical line segments have $z=1$, more typical for free-fermion systems. 

The critical exponents at and approaching the uniform case are something of a surprise. Another uncommon exponent comes by studying the gap as the multicritical point is approached.  Tuning to this point along the gapped $\alpha=\beta$ line gives a gap vanishing as $(\gamma-\alpha)^{3/2}$, as follows from \eqref{Bgamma} and \eqref{disper}. The same scaling occurs when approaching the multicritical point from any gapped direction.  Namely, varying the couplings slightly away from $1$ while fixing $\alpha+\beta+\gamma=3$ gives $B(\pi)\approx 1+\delta B$, where
\begin{align}
(\delta B)^2 =\frac{1}{3}\big(\delta\alpha\delta\beta+(\delta\alpha)^2+(\delta\beta)^2\big)\ ,
\end{align}
with the right-hand side indeed always greater than zero for a non-vanishing variation. The gap to leading order in the variations is then
\begin{align}
\Delta^2 \propto \big(2\delta B-\delta\alpha-\delta\beta\big)\big(2\delta B+\delta\alpha\big)\big(2\delta B+\delta\beta\big)
=2(\delta B)^3-\delta\alpha\delta\beta(\delta\alpha+\delta\beta)\ .
\end{align}
It is straightforward to check that the only directions where the gap stays zero are along the three critical line segments $\delta\alpha=\delta\beta>0$, $\delta\alpha=-2\delta\beta>0$ and $\delta\beta=-2\delta\alpha>0$. 
Thus indeed the gap vanishes with the same exponent $3/2$ that appears in the dispersion relation.

\section{Conclusion}

The models described in this paper have the remarkable feature that they can be solved by constructing free-fermion raising and lowering operators, despite the fact that they have no Hamiltonian written in terms of local fermion bilinears. Thus although the spectrum is given by the canonical free-fermion form \eqref{ffE}, other free-fermion characteristics need not apply. For example, the local fermions $\psi_j$ cannot be obtained as a linear sum of the $\Psi_{\pm k}$, since there are more of the former than the latter. Thus the approach here does not give much insight into the correlation functions save for the critical exponents and gap derived above. 

It is thus difficult to guess even qualitatively some of the physics will be beyond the results here. For example, since the model is solvable for any couplings, it should be possible to understand the behaviour when they are random. Presumably there is an infinite-randomness critical point as in the analogous random Ising chain \cite{Fisher95}, but is not at all obvious what it will look like. It is not even immediately clear how to adapt the decimation procedure to this model, since the interactions are not nearest-neighbour.

Of course, the situation here is rather unusual, in that each level has an exponentially large degeneracy $\sim 2^{M/3}$, as described in section \ref{sec:spectrum}, even with random couplings. Imposing periodic boundary conditions splits these degeneracies, and there is no particular reason to expect that the resulting energies will have the same dynamical critical exponent. Indeed, numerics \cite{Sannomiya17b} and analytic arguments \cite{AA19} suggest that these excitations will have $\epsilon\propto p^3$. Since as shown in appendix \ref{app:Rmat} the model remains integrable with periodic boundary conditions, this energy should be computable, However, the lack of an obvious $U(1)$ symmetry makes using the Bethe ansatz much more difficult, and I have not been able to do this calculation. Someone stronger should do it.

Another interesting direction to explore would be a connection of this very non-standard integrable model with other integrable systems. One intriguing observation is that the dynamical critical exponent $z=3/2$ at the uniform point is identical to that of the famous $1+1$-dimensional KPZ equation \cite{Kardar86}. Perhaps this correspondence is a coincidence. However, the fact that correlators for KPZ \cite{Sasamoto05} and free-fermion models such as the Ising model \cite{Wu75} can be written as Fredholm determinants hints that it is not.

Less ambitiously, the techniques described here almost certainly can be applied to simplify the solution of the ``free-parafermion'' model  \cite{Fendley14,Baxter14,AuYang14}. Indeed, the transfer matrix there can also be decomposed into a product form like \eqref{Gdef}, the observation that made possible the much simpler analysis here (early versions of this paper pre-dating this observation are even more horrible than \cite{Fendley14}). 
A more intriguing possibility for application of these techniques is the Cooper-pair chain of \cite{Fendley06}, shown to be equivalent to another interesting chain \cite{deGier15} in \cite{Feher17}.  Like here, the spectrum with open boundary conditions is free-fermionic even though the Hamiltonian is not comprised of free-fermion bilinears. Also like here, the Cooper-pair model is integrable and supersymmetric, with a underlying large non-Abelian symmetry algebra extending the supersymmetry.  The algebra of the Hamiltonian generators there is more complicated than \eqref{halg}, however, so the analysis may take a little more finesse.

\bigskip

I thank Dave Aasen and Jason Alicea for sharing their unpublished results, and Hosho Katsura for reminding me that $z=3/2$ in KPZ. This work was supported by EPSRC through grant EP/N01930X.

\appendix
\setcounter{equation}{0}
\renewcommand{\theequation}{\Alph{section}\arabic{equation}}

\section{The $R$ matrices}
\label{app:Rmat}


All the derivations in the main text followed from expressing the transfer matrix as the product \eqref{Gdef}, and then doing manipulations using the algebra \eqref{halg}. This product form is only valid for open boundary conditions, rendering the techniques fairly useless for periodic boundary conditions. In this section I first rewrite the transfer matrix in matrix-product operator, in the specific representation where $h_m$ is given by \eqref{hh}. I then find an ``$R$-matrix'' that makes it straightforward to show that transfer matrices at different values of the spectral parameter commute with either open or periodic boundary conditions.

The transfer matrix is a sum over products of $h_m=b_m\sigma^z_m\sigma^z_{m+1}\sigma^x_{m+2}$. For ease of notation, I give the derivation for the uniform case $b_m=1$, and then explain the generalisation to couplings varying in space. In order to define the appropriate Yang-Baxter equation, it is useful to rewrite the transfer matrix as a matrix-product operator, or in more traditional statistical-mechanical language, as a vertex model. The matrix-product operator (MPO) is of the form
\begin{align} 
T_{bc}(u)= \sum_{\{a_m\},\{t_m\}} A^{a_1}_{bt_1}A^{a_2}_{t_1t_2} A^{a_3}_{t_2t_3}\dots A^{a_L}_{t_{L-1}c}
\,\sigma^{a_1}_1\sigma^{a_2}_2\sigma^{a_3}_3\dots \sigma^{a_L}_{L}\ .
\label{TMPO} 
\end{align}
A nice picture for this MPO is
\begin{center}
\begin{picture}(205,45)
\put(9,29){$b$}
\put(0,26){\line(1,0){216}}
\put(18,-1){$\sigma^{a_1}_1$}
\put(24,10){\line(0,1){32}}
\put(33,30){$t_1$}
\put(48,10){\line(0,1){32}}
\put(42,-1){$\sigma^{a_2}_2$}
\put(72,10){\line(0,1){32}}
\put(57,30){$t_2$}
\put(66,-1){$\sigma^{a_3}_3$}
\put(96,10){\line(0,1){32}}
\put(120,10){\line(0,1){32}}
\put(144,10){\line(0,1){32}}
\put(168,10){\line(0,1){32}}
\put(192,10){\line(0,1){32}}
\put(124,0){$\cdots$}
\put(124,30){$\cdots$}
\put(183,-1){$\sigma^{a_{L}}_L$}
\put(170,30){$t_{L-1}$}
\put(201,29){$c$}
\end{picture}
\end{center}
Each vertical line corresponds to acting with the operator $\sigma^{a_m}_m$ on the Hilbert space, where $a_m=0,x$ or $z$, with $\sigma^0_m=1$ and the other two given by \eqref{sigmaam}. The horizonal line segments are labelled by the ``auxiliary'' or ``internal'' indices $t_m$.  Each crossing corresponds to the tensor $A^{a_m}_{t_{m-1}t_m}$ that gives the appropriate weighting for the corresponding term in the transfer matrix. 

Finding the appropriate elements of this tensor is easy given the simple form of the transfer matrix, where all the $h_{m_r}$ in a given term must commute with each other. The auxiliary indices run over the same three channels $0,x$ and $z$ that the $a_m$ do. Then the only non-vanishing elements of the tensor are $A_{00}^0$, $A^z_{0z}$, $A_{zx}^z$ and $A_{x0}^x$, and 
\begin{align}
A^0_{00}=1 \ ,\qquad\quad A^z_{0z}A^z_{zx}A^x_{x0}= -u\ ,
\label{Adef}
\end{align}
are the only constraints needed. The transfer matrix is then found by doing the sum in \eqref{TMPO} over all $a_m,t_m =0,x$ and $z$, with the end indices labelled by $b$ and $c$. For periodic boundary conditions, the transfer matrix is given by 
\begin{align}
T^{\rm per}(u)= \sum_{b=0,x,z} T_{bb}(u)\ 
\label{Tper}
\end{align}
with $L=M$, while for the open boundary conditions used in most of the paper it is
\begin{align}
T(u) = T_{00}
\label{Topen}
\end{align}
with $L=M+2$.
In the traditional language, this might be called an eight-vertex model, since there are eight possibilities for what happens at each vertex. However, this most definitely is not the famed eight-vertex model solved by Baxter \cite{Baxter82}, as there are three channels here in the auxiliary/internal space. 

The derivation of commuting transfer matrices follows from the  ``RTT'' relation. These involve a  $9\times 9$ matrix $R(u,u')$ whose matrix elements $R_{bc'}^{b'c}$ are labelled by letting $a,a',b$ and $b'$ each take the values $0, x$ and $z$. The art is to find an $R$ matrix that satisfies
\begin{align}
\sum_{d,d'}R_{bd'}^{b'd}(u,u') T_{d'c'}(u')T_{dc}(u)= \sum_{d,d'}T_{bd}(u)T_{b'd'}(u') R^{d'c}_{dc'}(u,u')\ ,
\label{RTT}
\end{align}
for all $b,b',c,c'$ and $u,u'$, along with 
\begin{align}
(R(u,u'))^{-1} = R(u',u)\ .
\label{Rinver}
\end{align}
If such an $R$ matrix exists, then it immediately follows that $[T^{\rm per}(u),T^{\rm per}(u')]=0$. For those with open boundary conditions to commute, the additional conditions
\begin{align}
R^{0b'}_{0b}(u,u')= \delta_{b0}\delta_{b'0}= R^{b'0}_{b0}(u,u')
\label{Ropen}
\end{align}
must be satisfied as well.

The beauty of the $R$ matrix approach is that the RTT relation can be reduced to a local one by  rewriting the transfer matrix as an MPO/vertex model as in \eqref{TMPO}. 
To find the matrix $R$, one needs to solve
\begin{align}
\sum_{d,d'}R_{bd'}^{b'd}(u,u') A'_{d'c'}A_{dc}= \sum_{d,d'}A_{bd}A'_{b'd'} R_{d'c}^{dc'}(u,u')\ ,
\label{RAA}
\end{align}
where the non-commuting operators $A$ and $A'$ are defined by 
\[
A_{cd}\equiv \sum_{a=x,y,z} A^{a}_{cd}(u)\sigma^a\qquad
 A_{cd}'\equiv \sum_{a=x,y,z} A^{a}_{cd}(u')\sigma^a
\]
with the tensor elements $A^{a}_{cd}(u)$ defined in \eqref{Adef}. Repeatedly applying \eqref{RAA} allows the $R$ matrix to be commuted from one end to the other, yielding \eqref{RTT}. 

Since there are 81 matrix equations in \eqref{RAA}, the system is wildly overconstrained. It is thus straightforward to work through and find what I believe is the unique solution for $R$, up to an overall constant and gauge choices. The $9\times 9$ matrix breaks into three $1\times 1$ blocks with
\begin{align}
R_{00}^{00}=R_{xx}^{xx}=R_{zz}^{zz}=1 \ ,
\end{align}
and three $2\times 2$ blocks
\begin{align}
\begin{pmatrix}
R_{dc}^{cd}& R^{cc}_{dd}\cr
R_{cc}^{dd}& R^{cd}_{dc}
\end{pmatrix}
\label{Rblock}
\end{align}
with $d\ne c$. All other matrix elements of $R$ vanish. The solution is easiest to display by making the gauge choice 
\begin{align}
A_{0z}^z(u)=A_{x0}^x(u)=\sqrt{u}\ ,\quad A_{zx}^z(u)=-1\ ,
\label{Agauge}
\end{align}
for $u$ positive.  
Then \eqref{RAA} and \eqref{Rinver} require
\begin{align}
&R_{x0}^{0x}=-R_{0x}^{x0}=R_{z0}^{0z}=-R_{0z}^{z0}=R_{zx}^{xz}=-R_{xz}^{zx}=\frac{u-u'}{u+u'}\ ,\cr
R_{xx}^{00}=R_{00}^{xx}&=R_{zz}^{00}=R_{00}^{zz}= \frac{2\sqrt{uu'}}{u+u'}\ ,\qquad R_{zz}^{xx}=\frac{2u'}{u+u'}\ ,\qquad R_{xx}^{zz}=\frac{2u}{u+u'}\ .
\label{Rsol}
\end{align}
These elements satisfy (\ref{Ropen}) as well, so the model is integrable for both periodic and open boundary conditions.

This $R$ matrix can be put in standard difference form by defining $u=\exp(2\theta)$ and $u'=\exp(2\theta')$, so that the blocks (\ref{Rblock}) for $d=0$ and $c=x$ or $z$ are
\begin{align}
\frac{1}{\cosh(\theta-\theta')}
\begin{pmatrix}
\sinh(\theta'-\theta)&1\cr
1&\sinh(\theta-\theta')
\end{pmatrix}\ ,
\end{align}
while the block for $d=z$ and $c=x$ multiplies the off-diagonal elements by $e^{\pm(\theta-\theta')}$. 
This $R$ matrix is a standard one, so perhaps there will be some way to write the full transfer matrix in some previously known form.

The transfer matrices commute even with spatially varying couplings. The MPO is defined simply by generalising \eqref{TMPO} to have position-dependent tensor elements $A^{a_m}_{t_{m-1}t_m}(u_m)$, with the gauge choice \eqref{Agauge} changed to
\begin{align}
A_{0z}^z(u_m)=b_m {u}\ ,\quad A_{x0}^x(u_m)=1\ ,\quad A_{zx}^z=-1\ ,
\label{Agauge2}
\end{align}
The proof is simply to check that \eqref{RAA} is satisfied for any gauge choice, and that in any gauge, the solution is invariant under rescaling $u$ and $u'$ by the same constant like \eqref{Rsol} is. Equivalently, it is a consequence of the $R$ matrix satisfying the difference property.

\section{Essential but technical proofs}

Here I provide the derivations of two essential results. One is that the commutation of the transfer matrices for open boundary conditions requires only the properties of the algebra \eqref{halg}, and so will apply to any model whose Hamiltonian and transfer matrices are generated by such $h_m$. The other is an identity essential in deriving the properties of the raising and lowering operators. In both cases, the product form of the transfer matrix is essential. 

I define some notation to simplify the equations. First of all,
\begin{align}  
c_m \equiv \cos\phi_m
\end{align}
while a rescaled Hamiltonian generator obeys
\begin{align}
h'_m\equiv\sin\phi_m\frac{h_m}{b_m}= -\frac{uh_m}{c_{m-1}c_{m-2}} \ .
\label{hprimedef}
\end{align}
Then
\[(h'_m)^2=\sin^2\phi_m\ =1-c_m^2,\qquad (g_m)^2=1+h'_m\ \] 
by using the definitions and the identity (\ref{gident1}).
Throughout the appendix, the $\pm u$ arguments are usually omitted, with $-u$ accounted for by defining $O^-\equiv O^{\rm T}(-u)$ for any operator. Then for example
\begin{align}
g_mg_m^{-}=c_m\ ,\qquad
g_m h_n = h_n g^{-}_m \ ,\qquad g^{}_m h^{}_{n}g_m^-=g_m^2h_{n}=(1+h'_m)h^{}_{n}\ ,
\end{align}
with $|m-n|=1,2$.


\subsection{Direct proof of commuting transfer matrices for open boundary conditions}

\label{app:CTMprod}

In this appendix I prove that the conserved charges $Q_M^{(s)}$ and the transfer matrices comprised of them commute among themselves. 
The key ingredient in the proof is to show that conjugating with $G$ implements a sort of duality. Namely, defining
\begin{align}
\widehat{h}_m \equiv h_m\frac{\cos\phi_{m+1}\cos\phi_{m+2}}{\cos\phi_{m-1}\cos\phi_{m-2}}\ ,
\label{htildedef}
\end{align}
means that 
\begin{align}
(\big(G_{M}(\pm u)\big)^{-1}\,Q^{(s)}_M G_{M}(\pm u)=\sum_{M\ge m_{r+1}>m_{r}+2} \widehat{h}_{m_1}\,\widehat{h}_{m_2}\,\dots\,\widehat{h}_{m_s}\ \equiv \widehat{Q}^{(s)}_M\ 
\label{GJJhat}
\end{align}
for any $u^2\ne u_k^2$. In this expression, one must set $\cos\phi_{M+1}=\cos\phi_{M+2}=1$ inside $\widehat{h}_M$ and $\widehat{h}_{M-1}$ along with the usual $\cos\phi_{0}=\cos\phi_{-1}=1$. 
This relation (\ref{GJJhat}) immediately leads to the commutation of transfer matrices at different values of the spectral parameter:
\begin{align}
Q^{(s)}&= G_{M}(u)  \widehat{Q}^{(s)} (G_{M}(u))^{-1}  =
 G_{M}(u) (G_{M}(- u))^{-1}Q^{(s)}  G_{M}(- u)(G_{M}(u))^{-1}   \cr
&= T_M(u) Q^{(s)}   (T_{M}(u))^{-1}\ .
\label{JJcommute}
\end{align}
Since this identity holds true for any $u\ne u_k$, $[Q^{(s)},Q^{(s')}]=0$ for all $s,s'$, and so transfer matrices at different values of the spectral parameter commute, proving \eqref{Tcommute}.


I first prove \eqref{GJJhat} for $s=1$ using induction. Let  $H_M$ and $\widehat{H}_M$ be the Hamiltonians with generators $h_m$ and $\widehat{h}_m$ respectively, and assume that
\[ (G_M)^{-1}   H_M G_M = \widehat{H}_M\ .\]
Then
\begin{align}
(G_{M+1})^{-1} H_{M+1}  G_{M+1}& =
\frac{1}{c_{M+1}}g^-_{M+1} \widehat{H}_M g^{}_{M+1} +
(G_{M+1})^{-1} h_{M+1} G_{M+1}\cr
&=\widehat{H}_{M-2} + \frac{g^-_{M+1}}{c_{M+1}}\left(\widehat{h}_{M-1}+ \widehat{h}_{M} 
+ \frac{g^-_Mg^-_{M-1}}{c_{M}c_{M-1}}  h_{M+1}\,  g_{M-1}^{}g^{}_M\right)g^{}_{M+1}\ .\qquad
\label{GHG}
\end{align}
A crucial point is that in this expression the $\widehat{h}_m$ are defined not including $c_{M+1}$ or $c_{M+2}$, e.g.\ $\widehat{h}_M=h_M/(c_{M-1}c_{M-2})$ here. 
The middle terms  in (\ref{GHG}) are simplified using 
\[{g^-_{M+1}}\left(\widehat{h}_{M-1}+ \widehat{h}_{M}\right) g^{}_{M+1} = \left(\widehat{h}_{M-1}+ \widehat{h}_{M}\right)
\left(1+h'_{M+1}\right)\ ,\]
while the final term can be simplified using
\begin{align}
g^-_{M+1}g^-_Mg^-_{M-1}h^{}_{M+1} g_{M-1}^{}g^{}_Mg^{}_{M+1}&=
g^-_{M+1}g^-_M h_{M+1}\big(1+h'_{M-1}\big) \, g^{}_Mg^{}_{M+1}\cr
&= g^-_{M+1}h_{M+1}\big(1+h_{M}'+c_{M}h'_{M-1}\big)\,g^{}_{M+1}\cr
&=c_{M+1} h_{M+1} + h_{M+1}\big(h_{M}'+c_{M}h'_{M-1}\big)\big(1+h_{M+1}'\big)\cr
&=c_{M+1} h_{M+1} - \big(h_{M}'+c_{M}h'_{M-1}\big)\big(h_{M+1}+b_{M+1}\sin\phi_{M+1}\big).\qquad
\label{ggg}
\end{align}
Using these simplifications in \eqref{GHG}, terms non-linear in the $h_m$ cancel because
\begin{align}
&\widehat{h}_M h'_{M+1}= \frac{h_M}{c_{M-1}c_{M-2}}\frac{-uh_{M+1}}{c_{M-1}c_M}
=h'_{M} \frac{h_{M+1}}{c_{M-1}c_M}\ ,\cr 
&\widehat{h}_{M-1} h'_{M+1} = h'_{M-1}c_M  \frac{h_{M+1}}{c_{M-1}c_M} 
= h'_{M-1} \frac{h_{M+1}}{c_{M-1}}\ .
\label{nlinident}
\end{align}
To simplify the linear terms in  \eqref{GHG}, note that
\[b_{M+1}\sin\phi_{M+1} = -u^{-1}(1-c^2_{M+1})c_Mc_{M-1}\ .\]
Therefore \eqref{GHG} becomes
\begin{align*}
G_{M+1}^{-1} H_{M+1}  G_{M+1} &= \widehat{H}_{M-2} 
+ \frac{1}{c_{M+1}c_{M-2}}\left(\frac{h_{M-1}}{c_{M-3}}+\frac{h_M}{c_{M-1}}\right)+\frac{h_{M+1}}{c_{M-1}c_M} 
+\frac{1-c^2_{M+1}}{uc_{M+1}}
 \big(h_{M}'+c_{M}h'_{M-1}\big)\cr
&= \widehat{H}_{M-2} 
+ \frac{h_{M-1}c_Mc_{M+1}}{c_{M-2}c_{M-3}}
+ \frac{h_{M}c_{M+1}}{c_{M-1}c_{M-2}} + \frac{h_{M+1}c_{M+1} }{c_{M-1}c_M}
 \end{align*}
using \eqref{hprimedef}.
The last three terms are precisely $\widehat{h}_{M-1}+\widehat{h}_{M}+\widehat{h}_{M+1}$, defined now so that they involve all $\cos\phi_{m}$ with $m\le M+1$ (but not $\cos\phi_{M+2}$ or  $\cos\phi_{M+3}$). Thus finally
\begin{align*}
(G_{M+1})^{-1} H_{M+1}  G_{M+1}  = \widehat{H}_{M+1}\ .
\end{align*}
Quite obviously,  \eqref{GJJhat} holds for $M=1,s=1$ because 
\[g_1^- h_1 g_1 =g_1^2 h_1=c_1h_1\ .\]
By induction, it then applies for $s=1$ and all $M$. 

The general duality relation \eqref{GJJhat} follows in a very similar manner. A crucial ingredient is the recursion relation for the  $Q_M^{(s)}$ found from \eqref{Trecur} and \eqref{TJdef}, namely
\begin{align}
Q_{M}^{(s)} = Q_{M-1}^{(s)} + h_{M} Q_{M-3}^{(s-1)}\ .
\end{align}
Assuming  \eqref{GJJhat} holds true for up to some fixed $M$,
\begin{align}
(G_{M+1})^{-1} Q^{(s)}_{M+1} G_{M+1}=
\frac{1}{c_{M+1}}g^-_{M+1} \widehat{Q}^{(s)}_M g^{}_{M+1} +
(G_{M+1})^{-1} h_{M+1}Q^{(s-1)}_{M-2} G_{M+1}
\label{GHG2}
\end{align}
where here $\widehat{h}_M$ and $\widehat{h}_{M-1}$ are defined without including $c_{M+1}$.
The first term in \eqref{GHG2} simplifies as in $s=1$ case: 
\[g^-_{M+1} \widehat{Q}^{(s)}_M g^{}_{M+1} =c_{M+1}\widehat{Q}^{(s)}_{M-2} + \left(\widehat{h}_{M}\widehat{Q}^{(s-1)}_{M-3} + \widehat{h}_{M-1}\widehat{Q}^{(s-1)}_{M-4}\right) \big(1+h'_{M+1}\big)\ .\]
This expression includes a slight abuse of notation: the $\widehat{h}_m$ inside the $\widehat{Q}_{M-2}^{(s)}$
keep all $c_m$ with $m\le M$. 
The second term in \eqref{GHG2}, however, requires more effort:
\begin{align}
G_{M+1}^{-1} h_{M+1}Q^{(s-1)}_{M-2} G_{M+1}=
\frac{g^-_{M+1}g^-_Mg^-_{M-1}}{c_{M+1}c_{M}c_{M-1}}  h_{M+1}
\Big(\widehat{Q}^{(s-1)}_{M-4}&+\frac{\widehat{h}_{M-2}}{c_{M}c_{M-1}} \widehat{Q}^{(s-2)}_{M-5}
+ \frac{\widehat{h}_{M-3}}{c_{M-1}}  \widehat{Q}^{(s-2)}_{M-6}\Big)
g_{M-1}^{}g^{}_Mg^{}_{M+1}
\label{secterm}
\end{align}
where again the $c_m$ with $m\le M$ are not set to 1.
Of the three terms in \eqref{secterm}, the first simplifies using \eqref{ggg}, the $g_m$ present commute through the middle term, and the last term is simplified via
\begin{align*}
\frac{g^-_{M+1}g^-_Mg^-_{M-1}}{c_{M+1}c_{M}c_{M-1}}  h_{M+1} \widehat{h}_{M-3}g_{M-1}^{}g^{}_Mg^{}_{M+1}&=
\frac{g^-_{M+1}}{c_{M+1}c_{M}}h_{M+1} \widehat{h}_{M-3} \big(1+h'_M\big)g^{}_{M+1}\cr
&=\frac{1}{c_{M}} h_{M+1} \widehat{h}_{M-3}+ \frac{1}{c_{M+1}c_{M}}h_{M+1}h'_M \widehat{h}_{M-3}(1+h'_{M+1})\ .
\end{align*}
Now I can group together all the terms in \eqref{secterm} multiplying $h_{M+1}(1+h'_{M+1})/(c_{M+1}c_Mc_{M-1})$:
\[ \big(h_{M}'+c_{M}h'_{M-1}\big)\widehat{Q}^{(s-1)}_{M-4}+h_M \widehat{h}_{M-3} \widehat{Q}^{(s-2)}_{M-6}
= h_{M}'\widehat{Q}^{(s-1)}_{M-3}+ c_{M}h'_{M-1}\widehat{Q}^{(s-1)}_{M-4}\ .\]
The other three terms in  \eqref{secterm} simplify to
\[\frac{h_{M+1}}{c_{M}c_{M-1}}\Big(\widehat{Q}^{(s-1)}_{M-4}-u\widehat{h}_{M-2}\widehat{Q}^{(s-2)}_{M-5}
-{u\widehat{h}_{M-3}} \widehat{Q}^{(s-2)}_{M-6}\Big) =
\frac{h_{M+1}}{c_{M}c_{M-1}}\widehat{Q}^{(s-1)}_{M-2}\ .\]
Combining all the terms in \eqref{GHG2}, the terms non-linear in the $h_m$ (not including those within the $Q$) all cancel as before using \eqref{nlinident}, leaving
\begin{align*}
(G_{M+1})^{-1} Q^{(s)}_{M+1} G_{M+1}&=
\widehat{Q}^{(s)}_{M-2} +\frac{1}{c_{M+1}} \left(\widehat{h}_{M}\widehat{Q}^{(s-1)}_{M-3} + \widehat{h}_{M-1}\widehat{Q}^{(s-1)}_{M-4}\right)+\frac{h_{M+1}}{c_{M}c_{M-1}}\widehat{Q}^{(s-1)}_{M-2}\cr
&\hskip.94in+\frac{1-c_{M+1}^2}{c_{M+1}}\Big(h'_{M}\widehat{Q}^{(s-1)}_{M-3}+ c_{M}h'_{M-1}\widehat{Q}^{(s-1)}_{M-4}\Big)\cr
&=\widehat{Q}^{(s)}_{M-2} +
\widehat{h}_{M}\widehat{Q}^{(s-1)}_{M-3}+ \widehat{h}_{M-1}\widehat{Q}^{(s-1)}_{M-4}
+\widehat{h}_{M+1}\widehat{Q}^{(s-1)}_{M-2}\cr
&=\widehat{Q}^{(s)}_{M+1}\ ,
\end{align*}
where now the $\widehat{h}_M$ and $\widehat{h}_{M-1}$ are defined including $c_{M+1}$. Thus indeed \eqref{GJJhat} holds for all $M$, and so transfer matrices commute.

\subsection{Proof of identity (\ref{TchiT})}
\label{app:TchiT}

An identity extremely useful in constructing the raising and lowering operators is given in  (\ref{TchiT}). Its proof is given here.
The first part in proving it is to rewrite the basic form of the raising and lowering operators as
\begin{align}
T_M\chi^{}_{M+1}T_M^- &=\sqrt{P_{M-1}}\,G_{M-1}\,(g_{M})^2\, \chi^{}_{M+1}\, (g_{M}^-)^2\, G^-_{M-1}\cr
&=\sqrt{P_{M-1}}\,
G_{M-1}\big(1+\sin^2\phi_{M}+ 2h'_{M}\big)  \chi^{}_{M+1}G^-_{M-1}\cr
&=2\sqrt{P_{M-1}}\,G_{M-1}\,(g_{M})^2\,\chi^{}_{M+1}\,G^-_{M-1} - \cos^2\phi_{M}\,{P^{}_{M-1}}\chi^{}_{M+1}\cr
&=2\sqrt{P_{M-1}}\,G_{M}\,\chi^{}_{M+1}\,G^-_{M} -P^{}_{M}\chi^{}_{M+1}\ .
\label{etagen}
\end{align}
With some more effort, the same type of simplification can be done for their commutator with $H$:
\begin{align*}
T_M\,h^{}_{M}\chi^{}_{M+1}T_M^- &=
\sqrt{P_{M-3}}\,G_{M-1}\,(g_{M})^2g_{M-1}g_{M-2}\,h^{}_{M}\chi^{}_{M+1}\, g^-_{M-2}g^-_{M-1}(g_{M}^-)^2\, G^-_{M-1}\cr
&=\sqrt{P_{M-3}}\,G_{M-1}\,(g_{M})^2\big(1+ h'^{}_{M-1}+c_{M-1}\,h'^{}_{M-2}\big)\,h^{}_{M}\chi^{}_{M+1} (g_{M}^-)^2\, G^-_{M-1}\cr
&=\sqrt{P_{M-3}}\,G_{M-1}\Big(1+\sin^2\phi_{M}+ 2h'_{M}
+ c^2_{M}\big(
h'_{M-1}+c_{M-1}\,h'^{}_{M-2}\big)\Big)\,h^{}_{M}\chi^{}_{M+1} \, G^-_{M-1}\cr
&=-2ub_{M}^2P_{M-3}\chi^{}_{M+1} + \sqrt{P_{M-3}}\,G_{M-2}\Big(\big(1+h'_{M-1}\big)
\big(2+c^2_{M}(h'^{}_{M-1}-1)\big)\cr
&\hskip2.9in +c^2_{M-1}c^2_{M}h'^{}_{M-2}\Big)\,h^{}_{M}\chi^{}_{M+1} \, G^-_{M-2}
\end{align*}
where to extract the term in front of the last equality, I used 
\[\sqrt{P_{M-3}}G_{M-1}\,h'_{M}h^{}_{M}G_{M-1}^-=\sqrt{P_{M-3}P_{M-1}}b_{M}\sin\phi_{M}=-u\,b_{M}^2\,P_{M-3}\ .\]
Then note that 
\[ (1+h'_{M-1})(2+c^2_{M}(h'^{}_{M-1}-1))=2(g_{M-1})^2-c^2_{M-1}c^2_{M}.\]
so that
\begin{align*}
T_M\,h^{}_{M}\chi_{M+1}T_M^- &= -2ub_{M}^2P_{M-3}\chi^{}_{M+1}+
2\sqrt{P_{M-3}}\,G_{M-1}\,h_{M}\chi^{}_{M+1} \, G^-_{M-1}\cr
&\qquad+\sqrt{P_{M-3}}c^2c_{M-1}c^2_{M}G_{M-2}(h'^{}_{M-2}-1)\,h^{}_{M}\chi^{}_{M+1} \, G^-_{M-2}\ .
\end{align*}
The last term is simply $-P_{M}h^{}_{M}\chi^{}_{M+1}$, because
\[ g_{M-2}(h'^{}_{M-2}-1)\,h^{}_{M}\chi^{}_{M+1} \, g^-_{M-2}=-c^2_{M-2}\,h^{}_{M}\chi^{}_{M+1}
\ , \]
while the middle term can be rewritten because
\[\sqrt{P_{M-3}}\,h_{M}^{}=-u^{-1}\sqrt{P_{M-1}}\,h'_{M}=u^{-1}\sqrt{P_{M-1}}(1-(g_{M})^2)\ .\]
Thus
\begin{align*}
T_M\,h^{}_{M}\chi_{M+1}^{}T_M^- =  \frac{2}{u}
\sqrt{P_{M-1}}\big(\sqrt{P_{M-1}}\chi^{}_{M+1}-G_M\chi^{}_{M+1}G_M^- \big) 
-2ub_{M}^2P_{M-3}\chi^{}_{M+1} -P_{M}h^{}_{M}\chi^{}_{M+1}.
\end{align*}
The recursion relation (\ref{Precur}) gives $P_{M-1}-u^2b_{M}^2P_{M-3}=P_M$, so finally
\begin{align}
u\,T_M\,h^{}_{M}\chi_{M+1}T_M^- &= -2\sqrt{P_{M-1}}\,G_{M}\,\chi^{}_{M+1}\, G^-_{M} 
+(2-uh_{M})P^{}_M\chi^{}_{M+1}\ .
\label{etagen2}
\end{align}
Adding \eqref{etagen2} to (\ref{etagen}) gives the desired identity \eqref{TchiT}:
\[T_M\,\big(1+uh^{}_{M}\big)\chi^{}_{M+1}T_M^- = \big(1-uh^{}_{M}\big)\chi^{}_{M+1}P_M\ .\]


\section{Derivation of dispersion relation}
\label{app:disper}

Here I provide the derivation of the dispersion relation for staggered couplings given and utilised in section \ref{sec:staggered} by deriving a matrix whose eigenvalues give the energy levels.

Imposing the staggering \eqref{stagg} and iterating the recursion relation \eqref{Precur} relates polynomials with indices differing by three:
\begin{align}\
P_{M} = \big(1-u^2(\alpha+\beta+\gamma)\big)P_{M-3} - u^4 (\alpha\beta+\beta\gamma+\alpha\gamma)P_{M-6}-u^6\alpha\beta\gamma\, P_{M-9}\ .
\label{Puniv}
\end{align}
In this form, the recursion relation is valid for $M\ge 9$, as long as $P_0=1$. Taking the number of generators $M$ to be a multiple of $3$ simplifies matters, as
the relation \eqref{Puniv} holds for $M=3$ and $M=6$ as well by defining $P_{-3}=P_{-6}=0$.  These recursion relations then can be recast as an eigenvalue relation by assembling the polynomials into a vector with entries
\begin{align}
v_s(\epsilon^2)= \epsilon^{-2s}P_{3s-3}(\epsilon^{-2})\ .
\end{align}
The recursion relation \eqref{Puniv} along with $P_{3S}(\epsilon_k^{-2})=0$ then gives
\begin{align}
\sum_{s'=1}^S\mathcal{R}_{ss'} v_{s'}(\epsilon_k^2) = \epsilon_k^2\, v_{s}(\epsilon_k^2)
\label{evalrelation}
\end{align}
for any $k,s=1\dots S$, with the entries of the $S\times S$ matrix $\mathcal{R}$ given by 
\[ \mathcal{R}_{ss'} = \delta_{s,s'+1} + (\alpha+\beta+\gamma)\delta_{s,s'} + (\alpha\beta+\beta\gamma+\alpha\gamma)\delta_{s,s'-1} +\alpha\beta\gamma\,\delta_{s,s'-2}
\ .\]
Crucially, these matrix elements depend only on $s-s'$, and only on the couplings. 

The energy levels  $\epsilon_k$ in the staggered case therefore can be computed simply by finding the eigenvalues of $\mathcal{R}$. The action of $\mathcal{R}$  on the ``plane-wave'' vector $v_s=\mu^{s}$ gives a solution of the eigenvalue relation \eqref{evalrelation} for $2<s< S$ if $\mu$ obeys 
\begin{align}
\epsilon^2\mu^2=\mu^3 + (\alpha +\beta+ \gamma)\mu^2 + (\alpha\beta+\beta\gamma+\alpha\gamma)\mu +\alpha\beta\gamma\ .
\label{cubicmu}
\end{align}
I label the three solutions of this cubic equation as $\mu_+$, $\mu_-$ and $\mu_0$, so that 
\begin{align}
(\mu+\alpha)(\mu+\beta)(\mu+\gamma)-\epsilon^2\mu^2=(\mu-\mu_+)(\mu-\mu_-)(\mu-\mu_0)\ .
\label{cubicmu2}
\end{align}
The eigenvectors of $\mathcal{R}$ are then linear combinations of the corresponding three plane-wave vectors:
\begin{align}
v_s = A_+ \mu_+^{s} + A_-\mu_-^{s}  + \mu_0^{s}\ .
\end{align}
Requiring that \eqref{evalrelation} be satisfied for $s=1$ and $s=2$ is equivalent to setting $P_{-3}=P_{-6}=0$, so the coefficients $A_\pm$ must obey
\begin{align}
A_+(\mu_+)^{-1}+A_-(\mu_-)^{-1}+(\mu_0)^{-1}=0\ ,\qquad\quad A_++A_-+1 = 0\ .
\label{Apm}
\end{align}
Requiring that \eqref{evalrelation} be satisfied for $s=S$ (i.e.\ $P_M(\epsilon_k^{-2})=0$) yields the ``standing wave'' condition
\begin{align}
A_+ (\mu_+)^{S+1} + A_- (\mu_-)^{S+1} +(\mu_0)^{S+1} =0\ .
\label{standingwave}
\end{align}

Combining the standing-wave condition \eqref{standingwave} with the cubic equation (\ref{cubicmu}) fixes the eigenvalues $\epsilon_k^2$ of $\mathcal{R}$.  To give a less implicit expression, it is convenient to reparametrise the solutions. Since all the coefficients of the cubic equation are real, there are two types of solutions. One type is where $\mu_+,\mu_-,\mu_0$ are all real, while the other type has one of them (say $\mu_0$) real, while the other two are a complex conjugate pair. For the standing-wave condition to apply in the limit of large $S$, two of the solutions must be of the same magnitude. Thus only the latter type of solutions can be used to build eigenvectors of $\mathcal{R}$, and so I parametrise
\begin{align}
\mu_+ = Be^{ip}\ , \qquad \mu_- = Be^{-ip}\ 
\end{align}
with $B$ and $p$ real. Furthermore, the roots of (\ref{cubicmu2}) must obey the $\epsilon$-independent relations \begin{align}\mu_+\mu_-\mu_0=-\alpha\beta\gamma\ ,\qquad \mu_+\mu_- + \mu_+\mu_0 + \mu_-\mu_0 = \alpha\beta+\alpha\gamma+\beta\gamma\ .
\label{cubic2}
\end{align}
Combining these two relations gives the cubic equation \eqref{Bp1} relating $B$ to $p$:
\begin{align}
B^2 - 2\alpha\beta\gamma B^{-1}\cos p=\alpha\beta+\alpha\gamma+\beta\gamma\ .
\label{Bp}
\end{align} 
The energy levels $\epsilon_k$ are then related to $B$ and $p$ by plugging any one of the roots into \eqref{cubicmu}. 
Using $\mu_0$ gives the expression \eqref{disper} that depends on $p$ only through $B$.

The standing-wave condition (\ref{standingwave}) determines the allowed values of $p$.  A solution for large system sizes exists when first two terms are the same order of magnitude and dominate the last. The first relation in \eqref{cubic2} requires that  $\mu_0 = -\alpha\beta\gamma/B^{2}$, so such solutions occur when $B>(\alpha\beta\gamma)^{1/3}$, i.e.\ \eqref{Bge}.
When \eqref{Bge} is obeyed, \eqref{standingwave} for large $S$ then reduces to
\[
e^{i2p(S+1)} =-\frac{A_-}{A_+}=\frac{\mu_+^{-1}-\mu_0^{-1}}{\mu_-^{-1}-\mu_0^{-1}}\ ,
\]
using (\ref{Apm}). Rewriting the roots in terms of $B$ and $p$ and simplifying using the cubic equation for $B$ then yields the quantization condition (\ref{pquant}). The right-hand side is indeed a phase, and in the limit of $S$ large, there are solutions for $0<p<\pi$. The standing-wave condition thus resembles the usual Bethe-ansatz quantization condition for the quasi-momentum. 


Gapless points can occur only when at least two of the couplings are equal. To see this, first note that taking the derivative of (\ref{Bp}) yields 
\begin{align}
\left(B^3+\alpha\beta\gamma\cos p\right)\frac{dB}{dp}= 
-{\alpha\beta\gamma}B\sin p\ .
\label{dBdp}
\end{align}
The expression in parenthesis on the left-hand side is always positive for $p\ne\pi$ because of the lower bound on $B$. Thus $dB/dp<0$ for all $0\le p< \pi$, and so the minimum of $B$ occurs as $p\to\pi$, the point at which two roots of \eqref{cubicmu} coincide and obey $\mu_+=\mu_-=-B$. Taking the derivative of \eqref{disper} and using \eqref{Bp} yields
\begin{align}
\frac{d\epsilon^2}{dp} = 2\frac{dB}{dp}\left(\frac{B^3}{\alpha\beta\gamma}+\frac{\alpha\beta\gamma}{B^3}+2\cos p\right)
\end{align}
so that $d\epsilon^2/dp<0$ except at $p=\pi$. Since $\epsilon^2\ge 0$ as a consequence of $H$ being Hermitian, any gapless points therefore must occur at $p=\pi$. Combining this with \eqref{disper} means that any gapless points must have $B^2(\pi)$ equal to any (or all) of $\alpha\beta$, $\beta\gamma$ or $\alpha\gamma$. A little algebra shows that \eqref{Bp} allows for $B^2(\pi)=\alpha\beta$ only if $\alpha=\beta$.  

Explicit expressions for the dispersion for $\alpha=\beta=1$ and $p$ near $\pi$ are found by computing $d^2B/dp^2$ using (\ref{dBdp}) and \eqref{Bgamma}, which for $\gamma\ne 1$ gives
\[
B \approx B_\gamma +  \frac{\gamma}{2(B^3_\gamma-\gamma)}(\pi-p)^2\ ,\quad\quad p\hbox{ near }\pi\ .
\]
A little algebra then yields
\begin{align}
\epsilon_\gamma(p)\approx 
\begin{cases}
({1-\gamma})^{1/2}|\pi-p|\ , \qquad\quad &\gamma<1\ ,\cr
\Delta_\gamma + C_\gamma (\pi-p)^2&\gamma>1\ ,
\end{cases}
\end{align}
where $C_\gamma$ a  positive constant and the gap $2\Delta_\gamma$ for $\gamma>1$ is
\begin{align}
2\Delta_\gamma = \gamma^{-1/2}(B_\gamma-1)(B^2_\gamma-1)^{1/2}\ .
\end{align}

\AtBeginEnvironment{thebibliography}{\linespread{.7}\selectfont}
\bibliography{newsusy}
\bibliographystyle{apsrev4-1}

\end{document}